\documentclass[preprint,aps,prx,showpacs,superscriptaddress,nofootinbib]{revtex4-1}
\usepackage{amssymb,amsbsy, amsmath}
\usepackage[dvipsnames,usenames]{color}
\usepackage{graphicx}
\usepackage{rotating}
\usepackage{dcolumn}
\usepackage[normalem]{ulem} 
\usepackage{xr}
\usepackage{xr-hyper}
\usepackage{hyperref}
\usepackage{nameref}
\usepackage{natbib}

\usepackage{zref-xr}
\zxrsetup{toltxlabel}

\usepackage{setspace}

\begin{document}

\title{Uncovering the socioeconomic facets of human mobility}

\author{Hugo Barbosa}
\affiliation{Department of Computer Science, University of Exeter, Exeter, UK}
\author{Surendra Hazarie}
\affiliation{Department of Physics \& Astronomy, University of Rochester, Rochester, NY, 14627, USA.} 
\author{Brian Dickinson}
\affiliation{Department of Computer Science, University of Rochester, Rochester, NY, 14627, USA.}
\author{Aleix Bassolas}
\affiliation{School of Mathematical Sciences, Queen Mary University of London, Mile End Road, E1 4NS, London, UK}
\author{Adam Frank}
\affiliation{Department of Physics \& Astronomy, University of Rochester, Rochester, NY, 14627, USA.} 
\author{Henry Kautz}
\affiliation{Department of Computer Science, University of Rochester, Rochester, NY, 14627, USA.}
\author{Adam Sadilek}
\thanks{sadilekadam@google.com}
\affiliation{Google Inc., 1600 Amphitheatre Parkway, Mountain View, CA, 94043, USA}
\author{Jos\'e J. Ramasco}
\thanks{jramasco@ifisc.uib-csic.es}
\affiliation{Instituto de F\'isica Interdisciplinar y Sistemas Complejos IFISC (CSIC-UIB), Campus UIB, 07122, Palma de Mallorca, Spain}
\author{ Gourab Ghoshal}
\thanks{gghoshal@pas.rochester.edu}
\affiliation{Department of Physics \& Astronomy, University of Rochester, Rochester, NY, 14627, USA.}
\affiliation{Department of Computer Science, University of Rochester, Rochester, NY, 14627, USA.}

\begin{abstract}
\singlespacing
Given the rapid recent trend of urbanization, a better understanding of how urban infrastructure mediates socioeconomic interactions and economic systems is of vital importance. While the accessibility of location-enabled devices as well as large-scale datasets of human activities, has fueled significant advances in our understanding, there is little agreement on the linkage between socioeconomic status and its influence on movement patterns, in particular, the role of inequality. Here, we analyze a heavily aggregated and anonymized summary of global mobility and investigate the relationships between socioeconomic status and mobility across a hundred cities in the US and Brazil. We uncover two types of relationships, finding either a clear connection or little-to-no interdependencies. The former tend to be characterized by low levels of public transportation usage, inequitable access to basic amenities and services, and segregated clusters of communities in terms of income, with the latter class showing the opposite trends. Our findings provide useful lessons in designing urban habitats that serve the larger interests of all inhabitants irrespective of their economic status.
\end{abstract}

\maketitle

\section{Introduction}

The recent trend of rapid global urbanization~\cite{DESAUN2018} poses major economic, social and structural challenges to cities~\cite{Danan2015}, accentuated by the environmental, and population impacts of climate change~\cite{Ford2019,Bassolas2019}. Added to this, rising economic inequality jeopardizes the health and livelihoods of many urban residents. Indeed, the percentage of Americans living in middle-income neighborhoods has decreased from 65\% to 42\% in the last 40 years, while the inhabitants of neighborhoods at the lower and higher ends of the income spectrum have grown \cite{Bischoff2014}. Much of the urban growth has been driven by the less wealthy who have progressively moved from the rural areas to cities~\cite{Massey1996}. While the rise in economic inequality is a global phenomenon, different patterns are observed between countries and cultures. For instance, while low-income households are typically located in the outskirts in Paris, they live downtown in Detroit \cite{Brueckner1999}. Such patterns of economic segregation and inequalities are connected as well to racial segregation, urban decay~\cite{Andersen2002} and gentrification~\cite{Brueckner2009}. In fact, over the course of history, many policies and resulting infrastructure changes were put in place in ways that hurt minority communities and already vulnerable portions of cities. This has been thoroughly documented in \cite{moses} for the case of New York and similar instances have been occurring around the world. Understanding the factors behind this and developing public policy to alleviate these trends is therefore crucial to enhance social mobility and economic progress, as well as positively impact the health of citizens~\cite{Ostendorf2001,Musterd2006,Eagle2010, Lobmayer2002}. 

Residential segregation is only a partial view of the whole picture, given that cities are a product of several interconnected systems. City infrastructure and dynamics are significantly connected to their underlying social systems, in particular influencing development and productivity indicators~\cite{Bettencourt2013,Pan_2013,Youn2016, Lee_2017, Kirkley_2018}. However, these mechanisms affect cities of different sizes in particular ways. On the one hand, technological advances are more rewarding to larger cities~\cite{henderson2010cities}, while urban democratization tends to be more beneficial for smaller cities~\cite{Henderson2007}. Underlying these intra-city mechanics are the people, and their mobility patterns~\cite{Barbosa2018}.
 
 \begin{figure*}[t!]
	\centering

		 \includegraphics[width =\textwidth]{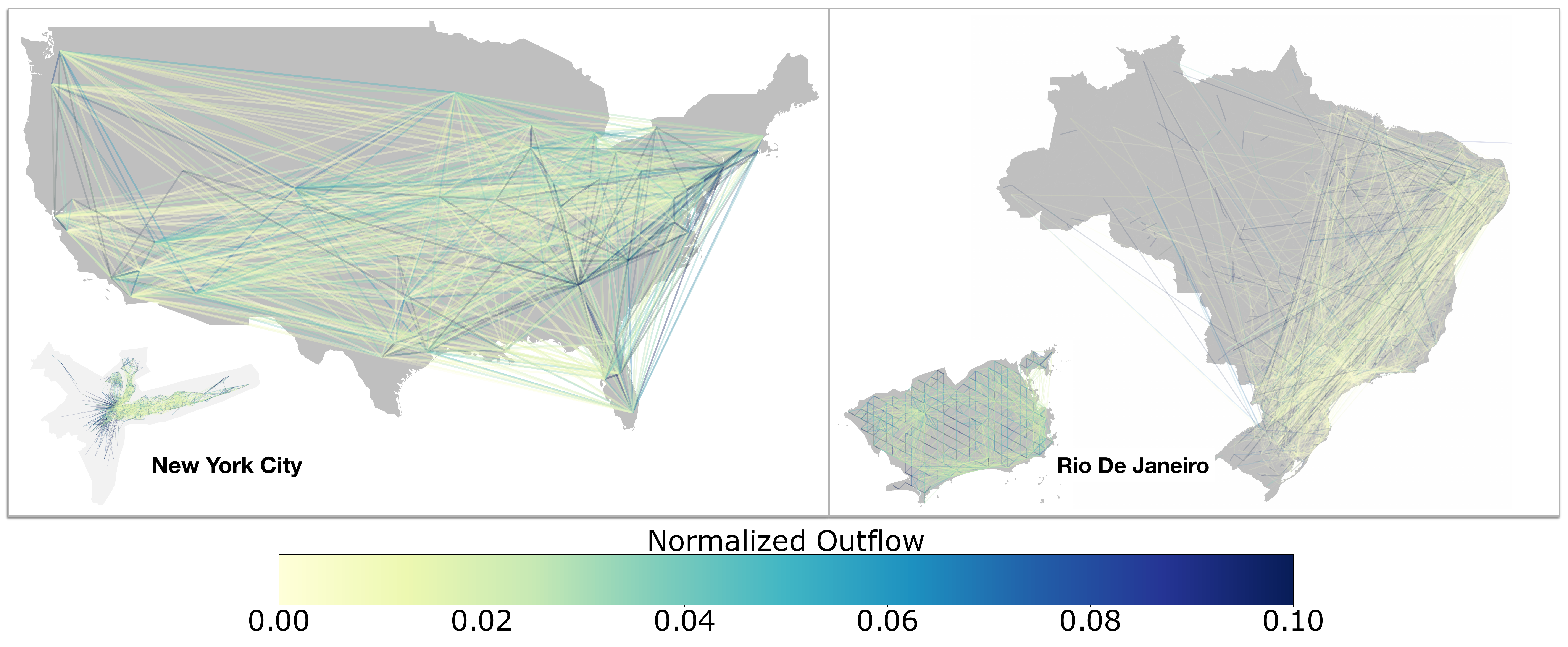}
	
	\caption{\textbf{Visualization of mobility data}. Flows at the country level for the US (left-panel), with New York City shown in the inset. Flows at the country level for Brazil (right-panel), with Rio shown in the inset.}
	\label{fig:flow_map}
\end{figure*}
 
Indeed, the increasing accessibility of location-enabled devices and large-scale datasets of human activities, such as credit-card purchases, taxi rides and mobile phone usage are fueling significant advances in our understanding of human mobility behavior~\cite{Yuan2012,Zhan2013,Lenormand2015,Wang2015,Luo2016,Louail2017,Gonzalez2008a,Di2018}. Yet, the connection of these observed mobility patterns with socioeconomic status is surprisingly unclear and in many cases contradictory.  For instance, while mobility patterns across different socioeconomic classes exhibit very similar characteristics in Boston and Singapore~\cite{Xu2018}, a similar study in Louisville, KY, revealed that low-income residents tended to travel further on average than those in affluent ones~\cite{Shelton2015}. On the other hand, analysis of cell phone data from an emerging Latin American economy revealed that wealthier citizens traveled to more locations and longer distances~\cite{FriasMartinez2012}. An analysis of mobility in Bogot\'a, Colombia found the most mobile population to be the upper-middle class instead of the wealthiest~\cite{Lotero2016}. Employment status played a role in movement patterns in both Riyadh and Spain, where the unemployed tended to travel less and spend more time at home~\cite{llorente2015,Rosenbaum2016a}. Two studies in French municipalities found a connection between the diversity of location visits with income, but no connection to the distance traveled~\cite{Pappalardo2015,Gabrielli2016}. These bewildering variety of results from studies around the globe indicates that that the observed diversity in trends are the result of a complex interplay between urban infrastructure and socioeconomic processes.  However,  each of these studies were conducted in a relatively small set of cities, different underlying datasets, and different methodologies. Greater clarity and insight may be uncovered by conducting the studies at scale, globally, and with a uniform methodology. 

To that effect, here, we explore the multiple facets connecting human travel behavior in a city to its socioeconomic landscape, through a country-wide analysis encompassing one hundred cities in the United States and Brazil. The two countries were selected due to their similarity in total population and the number of large cities, with one being a developed economy, while the other an emerging economy with markedly different socioeconomic characteristics. We find differences within and between the two countries, with two distinct classes of cities; one where there is a strong connection between the socioeconomic status of the residents and their movement patterns, and another class where there appears little-to-no connection. The latter class of cities are characterized by wider use of public transportation, equitable access to amenities and services as well as greater mixing among neighborhoods in terms of income profiles. The former class of cities show the opposite trends.
We conclude with a discussion of the implications of our findings, including possible policy directions as it relates to urban planning.

\section{Data description}

\subsection{Mobility data}
The Google Aggregated Mobility Research Dataset contains anonymized mobility flows aggregated over users who have turned on the Location History setting, which is off by default. This is similar to the data used to show how busy certain types of places are in Google Maps, helping identify, for instance, when a local business tends to be the most crowded. The dataset aggregates flows of people between regions, specifically ZIP Code Tabulation Areas (ZCTAs) for the United States, and the comparable census Weighting Area (WA) for Brazil, both of which we generically refer to as \textit{sub-areas} (SAs). 
 
To produce this dataset, machine learning is applied to logs data to automatically segment it into semantic trips \cite{Bassolas2019}. To provide strong privacy guarantees, all trips were anonymized and aggregated using a differentially private mechanism to aggregate the flows \cite{48778}. This research is done on the resulting heavily aggregated and differentially private data. No individual user data was ever manually inspected, only heavily aggregated flows of large populations were handled. As a result, the data is represented as a flow matrix $\mathbf{T}$ whose elements $T_{ij}$ correspond to annualized out-flows from location $i$ to $j$. For the purposes of our analysis we consider the 50 largest cities in the United States and analogously in Brazil ranked according to their population. In Fig.~\ref{fig:flow_map} we show the resulting mobility network for the United States and Brazil, with New York and Rio shown as inset. The nodes correspond to cities, and the edges are weighted normalized flows between the various locations. The full list of cities along with their populations are shown in Supplementary Tables S1 and S2. 

\subsection{Socioeconomic Indicators}

To represent socioeconomic status we collected data from the most recent 2016 five-year American
Community Survey (ACS) for the United States~\cite{census2016}, and analogously the 2010 decennial census for Brazil~\cite{censusBR2010}. Among the metrics collected were age, race, level of education, population, median income, sex composition, among others. A representative sample of the data, aggregated at the level of ZCTA's is shown for Rochester, NY in Supplementary Table S3, and at the level of WA's for Campinas in Supplementary Table S4. The choice of resolution is driven by the fact that both geographical levels are designated by the respective census authorities in such a way that their underlying populations are as socioeconomically homogeneous as possible. 

In Supplementary Figs. S1 and S2, we show correlation matrices displaying the Spearman rank correlation between a selection of indicators for both countries, finding that household median income has the strongest correlation with a number of other measures. Given that income is an indicator that is queried in most sociodemographic surveys throughout the globe and is the most direct way to estimate the economic capacities of different groups to afford mobility costs, it serves as a parsimonious measure. Consequently, without loss of generality and for the purposes of simplicity, in what follows, we use household median income as a low-dimensional proxy for socioeconomic status. 
We make use of two different income metrics. The first, $ID_{q}$, refers to the median income quintile of a given SA. When using this metric we define \textit{high} and \textit{low} income sub-areas as the top and bottom median income quintiles within their city. This metric is useful for simply categorizing the income of different sub-areas in a city. Our second metric, $ID_{f}$, is based on separate census-defined income breaks. In this case income is represented as a probability vector of the percentage of residents within each nationally defined income bracket and allows for more fine-grained analysis of income distribution. For more details on the metrics see Supplementary Section S1.

\subsection{Amenities}

Mobility patterns of urban residents are not only influenced by considerations of cost or income constraints, but also by the number and diversity of amenities and services accessible to them. To collect this information, we query the OpenStreetMap (OSM) database~\cite{osm}, that contains geo-referenced information for a broad array of urban amenities, including schools, banks, libraries, groceries and universities, to name a few. The OSM data contains 691 different types of urban facilities belonging to  eight main classes: healthcare (e.g., hospital and pharmacy), sustenance (e.g., restaurant and cafe), financial (e.g., bank and ATM), education (e.g., library and university), art-culture (e.g., arts center and theater), entertainment (e.g., cinema and nightclub), transportation (bicycle parking and bus station) and others (e.g., police station and post office). 

For the purposes of our analysis, we consider basic amenities related to food, healthcare, education and finance. A representative list of such amenities is shown for four different cities in the United States in Supplementary Table S5. To calculate the distance between a particular zip code and an amenity we use the geodesic distance between the centroid of the zip code and the coordinates of the amenity, represented by its latitude and longitude. This metric is an estimation of the characteristic distance the average individual in the zip code will need to travel to reach the amenity. It ignores population distribution within the zip code, as well as travel restrictions such as buildings and terrain to simplify computation. Despite these limitations, this distance is sufficient for our purposes.

\section{Results}

\subsection{Mobility metrics}

We characterize the mobility patterns of an area based on two representative quantities. The first metric is the average travel distance, or weighted average out-flow length, which is intended to capture the mobility-related costs that are directly proportional to travel distances. For each location $i$ it can be computed as
\begin{equation}
M_i^{dist} = \frac{\sum_j T_{ij}d_{ij}} {\sum_{j \neq i}T_{ij}},
\label{eq:Td}
\end{equation}
where $T_{ij}$ is the number of trips going from an area $i$ to a different area $j$ and $d_{ij}$ is the geodesic distance between the centroid of the two areas. 
The second metric we consider is the total flow per capita originating from location $i$, computed as
\begin{equation}
    M_i^{freq} = \frac{\sum_j T_{ij}} {P_i},
\label{eq:Tf}
\end{equation}
where $P_i$ is the population of location $i$. The city-level values are then computed by summing over all locations $i$ within the city. The measure serves as a proxy for trip frequency which is a useful complement to trip distance, as it represents an activity rate. Activity patterns, which are connected to employment status~\cite{Rosenbaum2016a}, combined with average distance traveled, indicate the distance and frequency of travel for residents in a given location. These factors are potentially influenced by the socioeconomic resources of the residents as well as their proximity to opportunities, and enables targeted ways of understanding the influence of city infrastructure and services on its residents. 

\begin{figure}[t!]
	\centering
	\includegraphics[width=\linewidth]{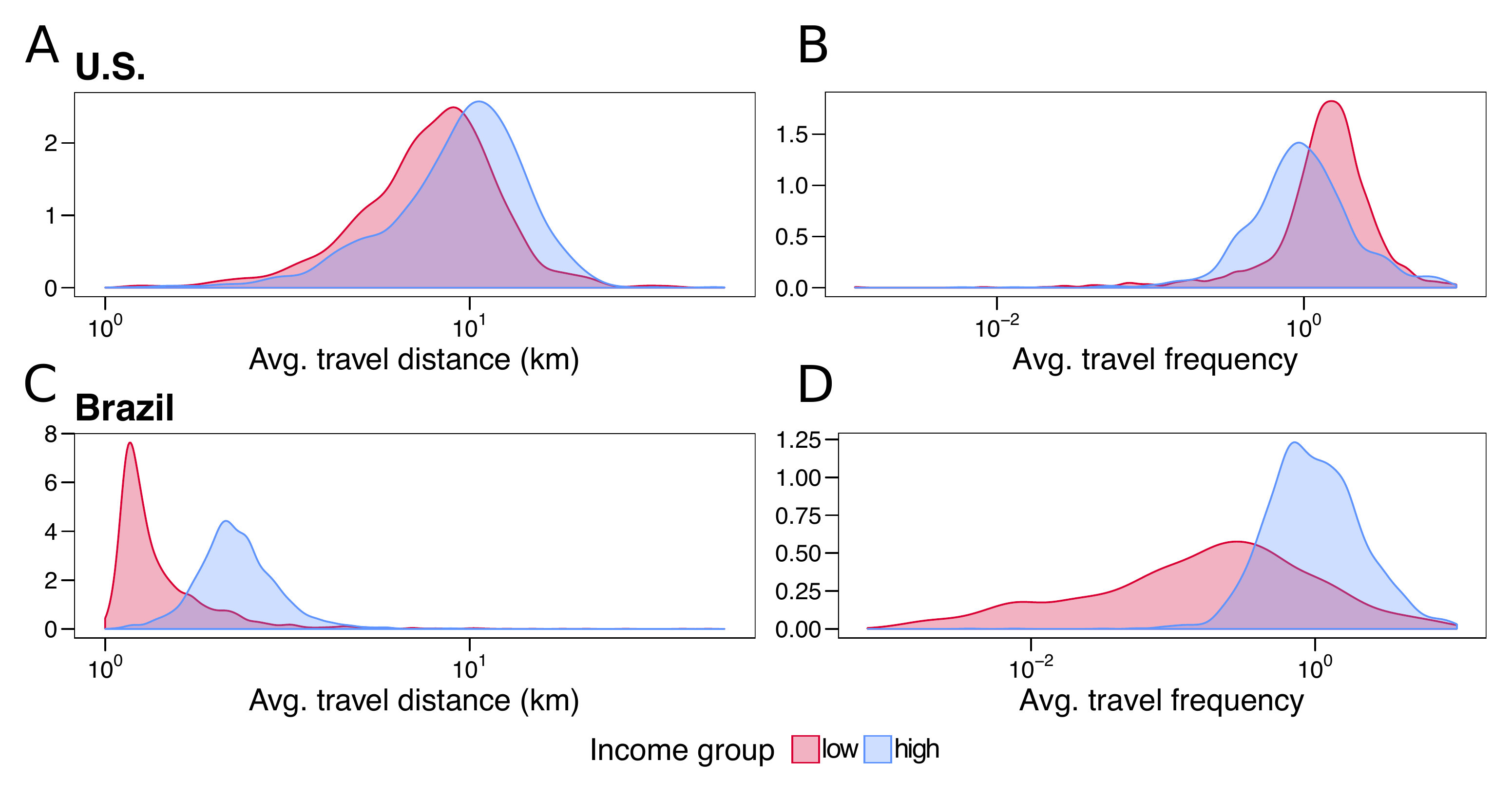}
	\caption{\textbf{Distribution of mobility metrics} Equations~\eqref{eq:Td} and~\eqref{eq:Tf} for the upper and lower 20\% residents in terms of income.  {\bf A} In the United States,  flows originating from low-income areas (red)  tend to be of slightly shorter length compared to the ones from the high-income regions (blue). \textbf{B} The opposite trend exists for travel frequency suggesting that wealthier residents travel less as compared to their poorer counterparts. \textbf{C.} Like in the United States, in Brazil, trips originating from low-income areas are significantly shorter in comparison with those from high-income zones. However, the discrepancy is more accentuated than in American cities. \textbf{D}, Unlike the United States, the frequency of trips is also much higher for wealthier residents in Brazil.}
	\label{fig:usbrmobilitykdeplotsv6labeled}
\end{figure}

We begin with a country-wide analysis of the mobility patterns of the lower and upper 20\% of SA's by median income. The choice of income range enables a first look at the trends for residents at the opposite end of the socioeconomic spectrum, that is those with the most and least potential constraints on their traveling behavior. In Fig.~\ref{fig:usbrmobilitykdeplotsv6labeled}{\bf A,B} we plot the probability density functions for $M_i^{dist}$ and $M_I^{freq}$ in the United States and in Fig.~\ref{fig:usbrmobilitykdeplotsv6labeled}{\bf C,D} for Brazil. In the US, we find that poor residents generally travel shorter distances but more often than the rich. Conversely, in Brazil, we see that mobility is dominated by the wealthier individuals, who travel more frequently and for longer distances, with the magnitude of separation in the mobility metrics between the different income groups being more pronounced in Brazil. However, for both countries, there is considerable width and overlap to the distributions, suggestive of a diversity in trends for individual cities. To check whether these observed features are due to biases in the data coverage based on income, in Fig. S3, we plot the mobility outflow as a function of the population for both income brackets in the two countries. We find a strong monotonic dependence, indicating that the flow provides equal representation of population irrespective of income. Indeed this is reflective of the fact that both over $90\%$ of households in the United States and Brazil own mobile phones and both countries rank among the top five in terms of smartphone usage\footnote{https://www.pewresearch.org/global/2016/02/22/smartphone-ownership-and-internet-usage-continues-to-climb-in-emerging-economies/}.

\subsection{Socio-mobility correlations}

\begin{figure*}[t!]
	\centering
	\includegraphics[width=0.8\textwidth]{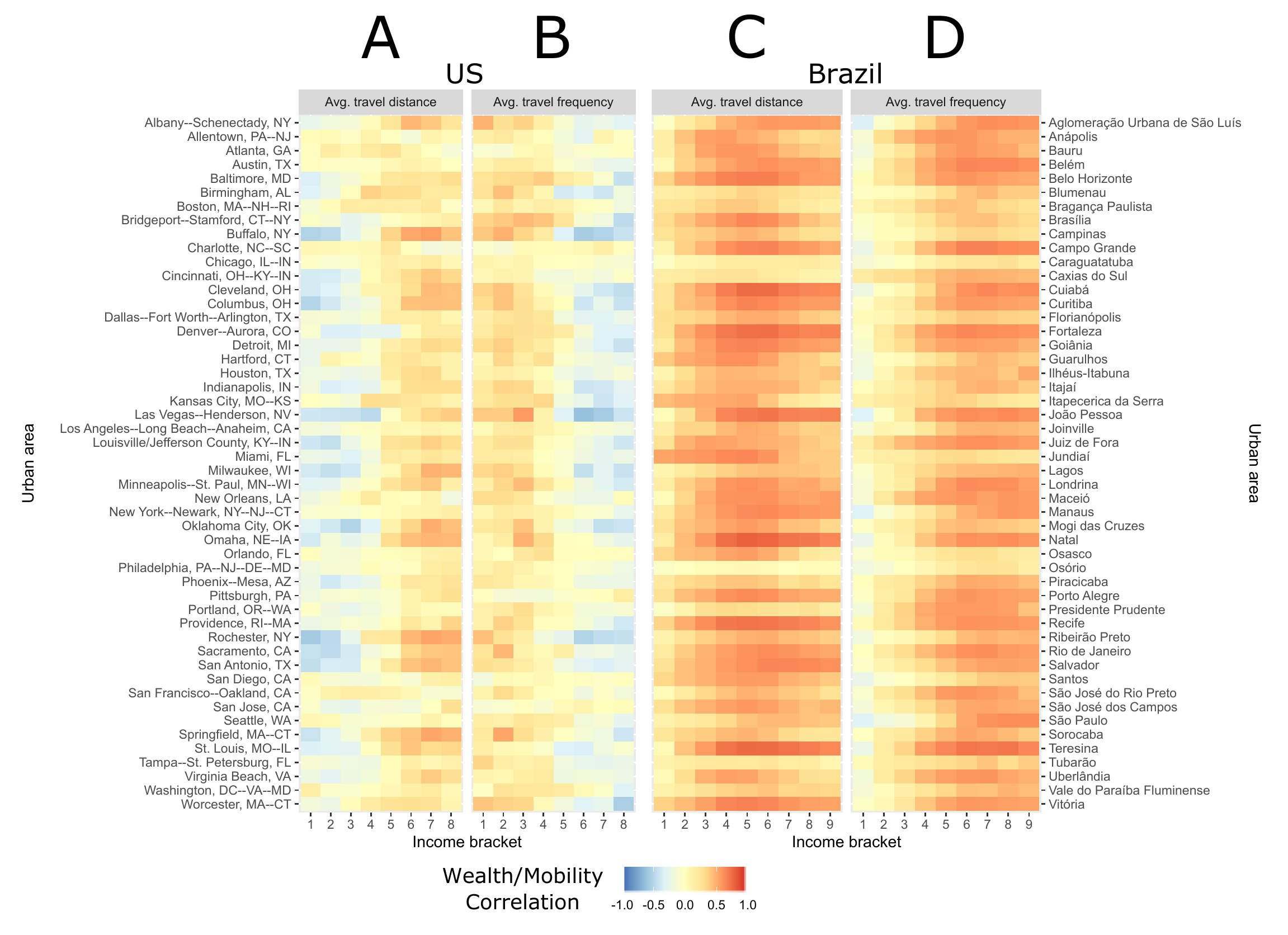}
	\caption{\textbf{Correlation between income and mobility.} For each income bracket (divided into eight buckets), Spearman correlation between the share of residents and the average travel distance, as well as average travel frequency for the United States (\textbf{A} and \textbf{B}) and Brazil (\textbf{C} and \textbf{D}). The colors indicate the level of correlation from dark-blue (-1) to dark orange (+1).}

\label{fig:mobility_income_correlations}
\end{figure*}

Given the observed country-level signal between the mobility metrics at the opposite ends of the income scale, we next conduct a more granular analysis across all census-determined income breaks $ID_{f}$, and at the level of individual cities.  Given an income distribution of $n$ income intervals, we construct for each urban area, $n$ income distribution vectors $\mathcal{I}$ of $m$ positions each $\mathcal{I}_{j_{1\ldots n}} = [f_{1}, f_{2},\ldots,f_{m}]$, where $m$ denotes the number of SAs in that city. Here, $f_m$ is the probability that an individual within the income interval $n$ lives in SA $m$. We compute the Spearman rank correlation coefficient between each $\mathcal{I}_{j}$ and the vectors for $M^{freq}$ and $M^{dist}$ for all $m$. For the results reported here, $n=8$ for the United States and $n=9$ for Brazil as determined by the respective censuses.

In Fig.~\ref{fig:mobility_income_correlations} we plot these correlations for both mobility metrics for all hundred cities. In the United States, we see two distinct correlation patterns for both mobility variables: cities with strong correlations (depicted as darker colors) or cities with weak correlations (lighter colors) across income brackets. For the lower-income brackets in cities like San Antonio and Sacramento there is a marked negative correlation with $M^{dist}$ (Fig.~\ref{fig:mobility_income_correlations}{\bf A}) and a positive correlation with $M^{freq}$ (Fig.~\ref{fig:mobility_income_correlations}{\bf B}), indicating that low income populations in those cities tend to travel over shorter distances but more frequently. Surprisingly, for these same cities, when we look at the high-income regions of the plot, we see an opposite relationship, with trips being longer (positive correlation with $M^{dist}$) and less frequent (negative correlation with $M^{freq}$), a pattern that is observed in many other cities. Conversely, the other remaining cities display little-to-no socio-mobility correlations. For example, in New York, Seattle, Chicago and Washington DC, the correlations are quite weak throughout the entire income range, indicating that income has little influence on how far or how frequently people travel.

The Brazilian cities, on the other hand display markedly distinct trends. While there are a few cities with relatively little correlations such as Caxias do Sul, Campinas and Blumenau, the majority of cities display high positive correlations across metrics and income ranges, with the strongest signal being in the middle-income range (Fig.~\ref{fig:mobility_income_correlations}{\bf C,D}). Whereas in the US, we see a mix of positive and negative correlations, in Brazil the correlations are by and large positive, except for the lowest income bracket being anti-correlated with travel frequency in a number of cities. The trends indicate a distinct mobility advantage for those higher up in the income scale both in terms of travel distance and frequency. 

\begin{figure*}[!t]
    \centering
    \includegraphics[width=0.8\linewidth]{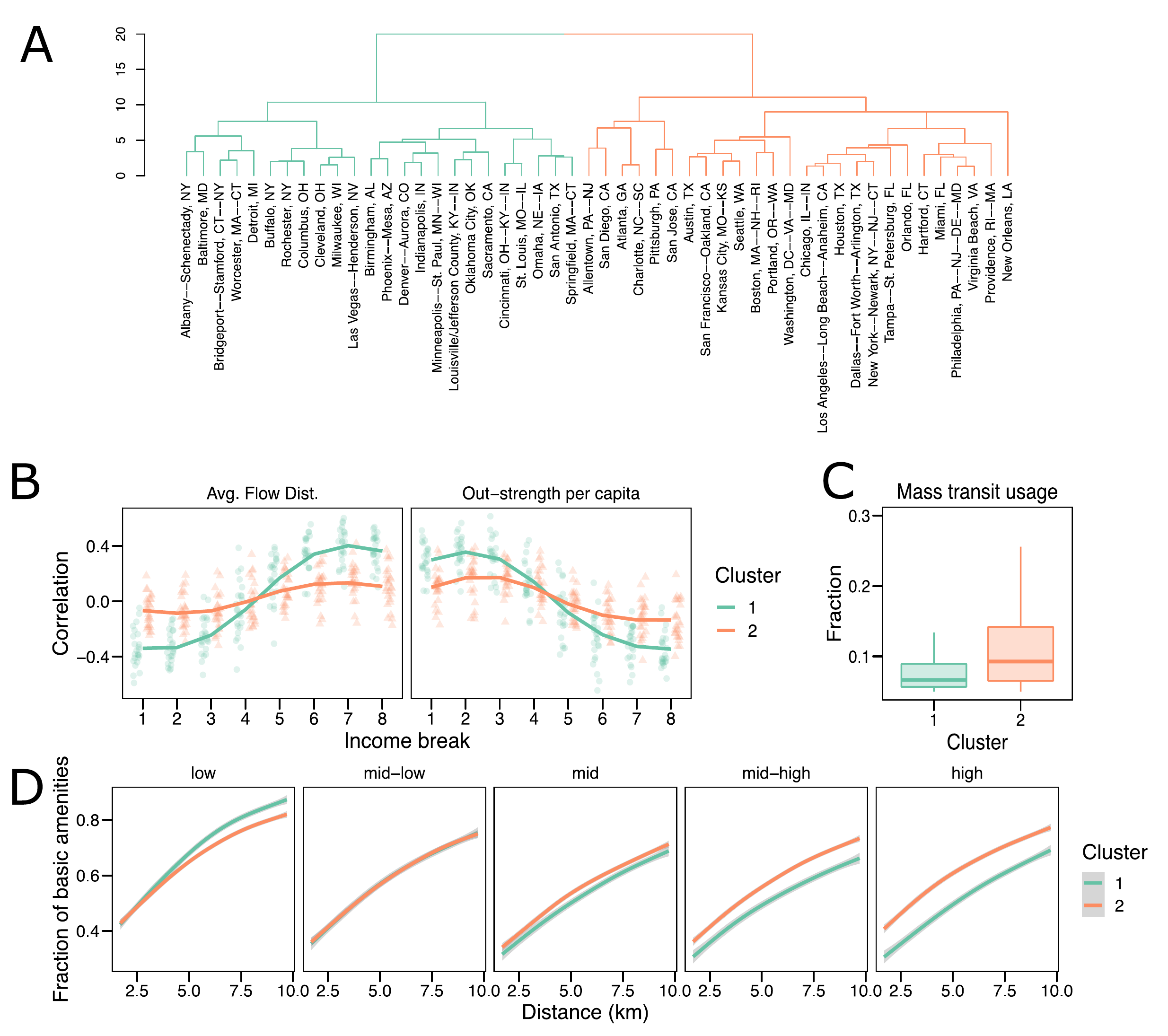}
    \caption{\textbf{Clustering analyses of cities in United States} {\bf A}  Dendogram obtained from a divisive hierarchical clustering method to partition cities into two different groups color-coded in teal and orange. \textbf{B} Spearman correlation between eight income breaks ($ID_{f}$), and the average flow distance and out-strength per capita. \textbf{C} Fraction of the population using mass transit for each of the clusters. \textbf{D} Fraction of basic amenities accessible as a function of distance for 5 income breaks ($ID_{q}$) for each of the clusters.}
    \label{fig:usa_clusters}
\end{figure*}

\subsection{City Clusters}
The results from Fig.~\ref{fig:mobility_income_correlations} suggest a multiplicity of patterns in terms of the connection between mobility and income in both countries. In order to check if cities can be classified into distinct groups according to these trends, we implement a clustering method to partition the cities according to their socio-mobility correlation patterns.  Using the Manhattan distance between the pairs of correlation values in each city, we perform a divisive hierarchical clustering~\cite{Geunoche_1991}, finding a clear partition of the cities in the United States into two different groups, color-coded in teal and orange as shown in Fig.~\ref{fig:usa_clusters}{\bf A}. Some interesting trends are immediately apparent. The largest seven cities ($\geq5$M inhabitants) are all in the orange cluster. On the other hand, the remaining cities are evenly distributed across both clusters, suggesting that population size is not a key factor in the observed partitioning (see Supplementary Section S4 for details on the clustering). 

In Fig. \ref{fig:usa_clusters}{\bf B} we plot the correlations across the 8 income breaks, separately for each cluster. A clear distinction emerges, whereby the teal cluster exhibits both stronger positive and stronger negative correlations, with lower income brackets traveling shorter distances but more often than the wealthy. This is in stark contrast to the orange cluster, where income and mobility variables exhibit relatively flat correlations across all the income breaks. These cluster assignments, quantitatively confirm the trends seen in  Fig.~\ref{fig:mobility_income_correlations}{\bf A}, and suggests the existence of two different classes of cities in the United States; those where travel patterns are determined by income, and those which are broadly independent of socioeconomic characteristics. 

We next investigate the factors behind this difference in correlations. A way to mitigate the disadvantages due to income is having efficient and cheap public transportation systems. In Figure \ref{fig:usa_clusters}{\bf C} we plot the percentage of individuals that use public transit in either cluster (extracted from the census), finding that a larger fraction of residents in the orange cluster use mass transit methods compare to those residing in the teal one. We note from a previous analysis~\cite{Bassolas2019}, that those cities that are in the orange cluster tend to be more hierarchical and centralized in terms of mobility-flows as compared to those in the teal cluster that are sprawled and decentralized. Hierarchical cities have higher levels of infrastructure for public transit, and the observed higher usage of public transit is not just a behavioral feature but also reflects wider availability. 

As mentioned earlier, another feature that might influence mobility patterns is the availability of amenities and services. Correspondingly, we check the typical distance residents of a particular location have to travel before they come across a number of basic amenities (food, healthcare, education and finance). For ease of display, instead of using all eight income breaks, we split the income groups into five categories (low, mid-low, mid, mid-high) and plot the fraction of amenities accessible up to 10km from the resident location.(See Supplementary Section S1 for details on the income-breaks). Using this data we analyze the distances different income groups, $ID_q$, would have to travel, in order to reach a certain percentage of basic amenities and plot our results in Fig.~\ref{fig:usa_clusters}{\bf D}. In the orange cluster, the accessibility of amenities is relatively stable across income groups, indicating that that in cities where low and high income groups move in similar ways, they tend to be at similar distances from basic amenities. This is either due to a more equitable distribution of resources, or the fact that both the wealthy and the less-so, tend to live in roughly the same areas. In contrast, within the teal cluster, we see an interesting trend, whereby, the distance traveled to reach the same fraction of amenities increases with income. In other words, the less wealthy live closer to basic amenities (at least in quantity, if not quality, a feature we investigate later), as compared to those with higher income. This is likely a reflection of the differences in urban organization between the orange and teal clusters, where in the latter, more sprawled configuration, lower income groups live in the inner-city (where a number of amenities are in close proximity), with the wealthier residents living in suburbs.  

In order to check whether our results are broadly stable across multiple datasets of human mobility, we next conduct a similar analysis on publicly available commuting data from the United States census bureau's LODES database~\cite{lodes}. Unlike the aggregate location history signals, this captures primarily work-home commuting trips. Nevertheless the aggregate mobility flows are strongly correlated with the LODES data up to around $10^3$ commuters as seen in Fig. S5.  Beyond this limit, the datasets diverge significantly indicating that the mobility data contains more information on non-commuting flows. With respect to the trends seen in Fig.~\ref{fig:usbrmobilitykdeplotsv6labeled}, the commuting data is in qualitative agreement in terms of the average distance traveled, but differs in terms of the trip-frequency (Fig. S6). Here, there is little difference between the income groups in terms of how often they travel. The key reason here is the way that work-home commutes are calculated from the LODES database. Irrespective of how many times an individual commutes to work over the year, it is counted as a single instance of the flow. Thus apart from not capturing non-commuting flows, the LODES data is limited in terms of accurately reflecting the true number of trips made between locations. 

With this in mind, we reproduce the clustering analysis conducted in Fig.~\ref{fig:usa_clusters} for the commuting data and plot the results in Fig. S7. The resulting clusters are shown in Fig. S7{\bf A} and a comparison to the divisions in Fig.~\ref{fig:usa_clusters} yields a Fowlkes-Mallows index score of 0.8~\cite{FM_1983}, indicating that the splits are by and large similar. Yet there are some notable differences; cities such as Portland, Seattle and Kansas City that were originally in the orange cluster are now found in the teal cluster. On the other hand, some cities in the teal cluster such as Detroit, Milwaukee and St. Louis move to the orange cluster. The correlation of the average distance traveled with $ID_{f}$, mirrors that seen for Fig.~\ref{fig:usa_clusters} although the differences between the two clusters are less pronounced (Fig. S7{\bf B}). On the other hand, the correlations with trip-frequency are markedly different, with the teal cluster now showing a flat trend, and the orange cluster showing a small positive trend with income. In terms of public transportation use in each cluster (Fig. S7{\bf C}) the trends are the same, largely due to the fact that the major metropolitan areas with high prevalence of public-transit-usage remain in the orange cluster for both datasets. A similar pattern emerges in terms of the distance to amenities, split by $ID_{q}$, with the orange cluster having little-to-no income dependence on amenity accessibility, whereas in the green cluster the higher income groups on average live further away from basic amenities (Fig. S7{\bf D}). Subtle differences exist in terms of the precise fraction of amenities accessed between the two datasets. Thus the primary difference between the location and commuting data, is the weaker or lack of correlations with mobility metrics in the latter due to the lower resolution in trip-frequency. This results in the switching of some cities between the clusters (some rather misleading, as in putting Detroit and NYC in the same cluster as we will show later). The qualitative features extracted from both datasets, however, are the same attesting to the robustness of the analysis. 

\begin{figure}[t!]
	\centering
	\includegraphics[width=0.7\linewidth]{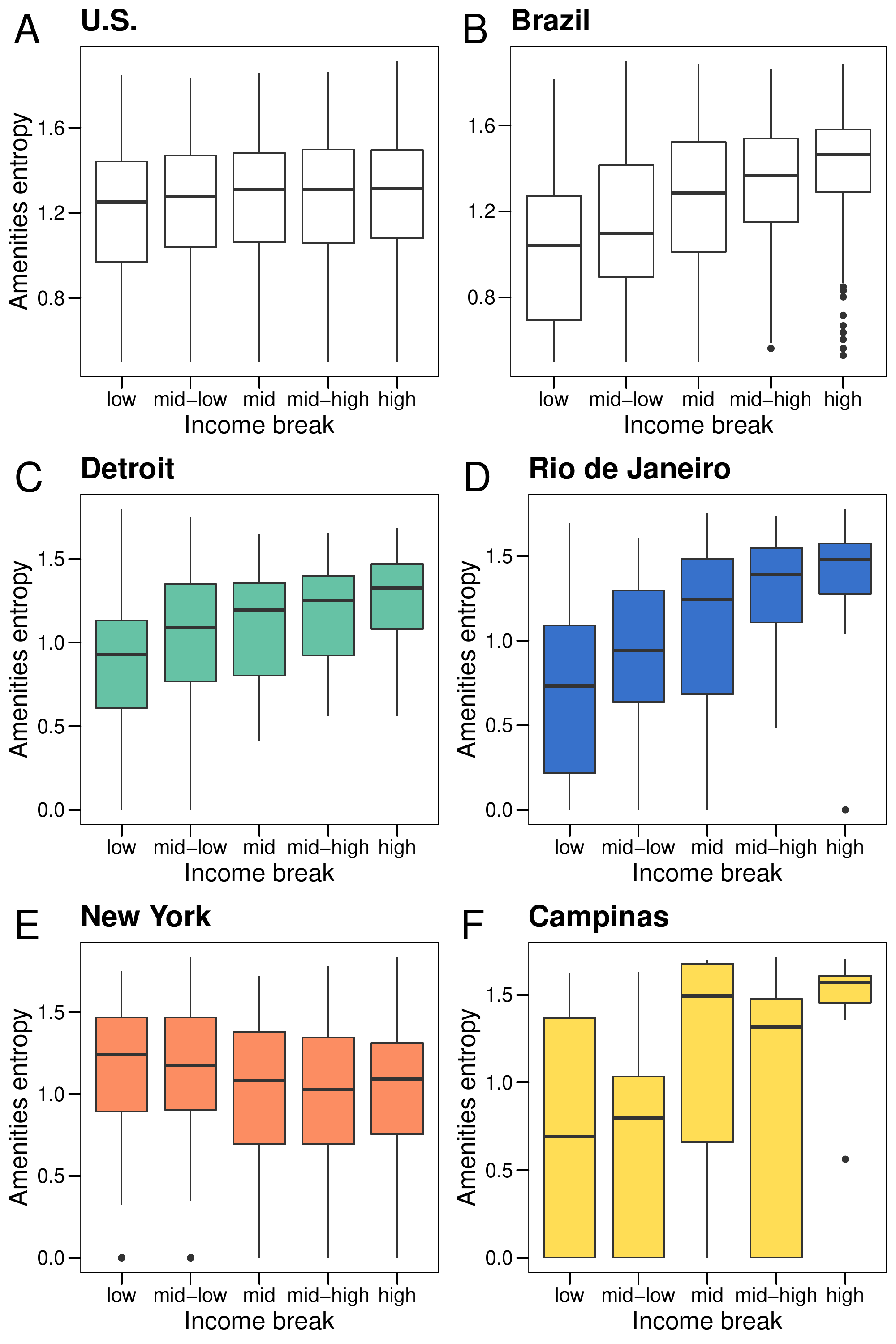}
	\caption{\textbf{Diversity of urban amenities per median income quintile.} Entropy of amenities in each of the sub-areas as a function of the median income in the United States (\textbf{A}) and Brazil (\textbf{B.}). Entropy of amenities in each of the sub-areas as a function of the income in Detroit (\textbf{C.}), a typical city of the teal cluster in the United States. Areas with higher median income tend also to have a higher diversity of amenities, a similar pattern seen for Rio de Janeiro (\textbf{D}) a typical city of the blue cluster in Brazil. In New York City, NY (\textbf{E}) (orange cluster), the trends are flat, and in Campinas (\textbf{F}) noisy, though higher-income areas on average, have higher amenity diversity.}
	\label{fig:amenitiesdiversitydetroitbostonriocampinas}
\end{figure}

Next, we analyze the cities in Brazil (Supplementary Figure S8{\bf A}), resulting in two clusters of roughly the same size,  color-coded yellow and blue. Unlike the more mixed patterns in the United States in terms of population size, nearly all the Brazilian cities in the blue cluster are the largest metropolitan areas and the capital cities of their respective states, with the exception of Lagos and S\~ao Jos\'e dos Campos. Among the cities in the yellow cluster, the only state capital is Florian\'opolis.
In Supplementary Fig. S8{\bf B}, we plot the correlation of the mobility metrics with $ID_{f}$ for each of the clusters, finding that the trends are quite different from that found in the United States. In both clusters there is an increase in correlation with both metrics as one goes up the the income ladder, with the strongest connection being in the middle income range. The correlations are always positive and stronger in magnitude than in the United States. In the yellow cluster while the lowest and highest income ranges travel similar distances, the wealthier travel much more frequently. In the blue cluster, the wealthy travel much longer distances and more frequently than the poor. Indeed, in both clusters the poor have similar movement patterns. 

Public transit usage (Supplementary Fig. S8{\bf C}) is moderately higher in the blue cluster (which also contains the largest cities in Brazil), but given the the similar trends seen for the mobility patterns for the lower income groups in both clusters, it does not play the same role as an equalizer as seen in the United States. 
In terms of access to the fraction of amenities, Supplementary Fig. S8{\bf D} indicates that for both clusters, access is skewed to the higher income ranges, who have to travel less to access services as compared to the poor. The main difference between the two clusters is that residents in the blue group, have an even higher advantage for the wealthy as compared to the yellow cluster.  In many ways all of the Brazilian cities are similar to the teal cluster in the United States, the difference being the correlations with income are even more pronounced, and the the wealthy have an advantage in both flavors of mobility metrics.

\subsection{Diversity of amenities}

In the teal cluster for the United States, we found that the lower income groups were more proximate to basic services as compared to those in the higher income range. Note that this is a measure only of the number of such basic amenities and provides no information on the quality of such services. While defining quality is difficult and can be probably only be extracted from targeted surveys, a reasonable proxy is to measure the mix of amenities. Indeed, for instance, lower income residents in inner-cities may have access to a number of grocery stores, or corner-shops but may be lacking in financial or health-care services. To determine, this we examine the ``diversity" of amenities for each each SA in the cities by measuring the entropy $H_l ^a$ in each SA $l$ (see Supplementary Section S6 for details of the calculation). Higher values of the entropy indicate a more homogenous distribution of the types of basic amenities, while lower values indicate that the distribution is dominated by a sub-type. 

\begin{figure*}[t!]
	\centering
	\includegraphics[width=\textwidth]{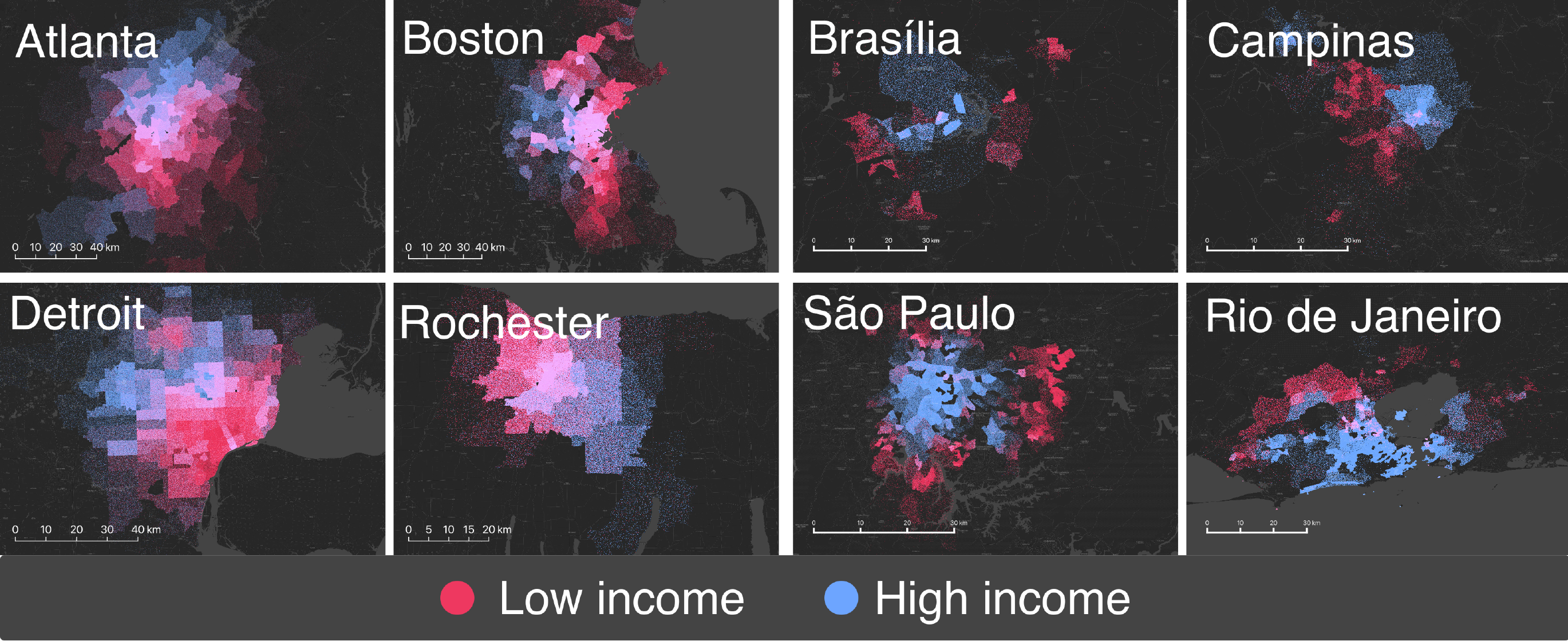}
	\caption{Socioeconomic in-flow heatmaps for selected cities in the US and Brazil. Colors represent the income level of the origin area that produced mobility flow into the plotted destination region. We see that cities such as Atlanta and Boston in the US experience considerable overlap between low and high income destinations, suggesting that income is not a defining factor in where people are able to live and go. In other cities like Detroit and Rochester, we see that the destinations of high and low income residents are relatively partitioned, suggesting that the particular land use decisions made in these cities have allowed lower income residents better access to amenities by living in the downtown areas. We see consistent separation between high and low income destinations in Brazil across all cities, suggesting that amenity access is generally shaped by residential location.}
	
	\label{fig:spatial_mixing}
\end{figure*}

In Fig.~\ref{fig:amenitiesdiversitydetroitbostonriocampinas}{\bf A,B} we plot $H_l ^a$ according to the income buckets $ID_q$ at the country-level for the United States and Brazil. For the United States, the entropy is relatively flat across income, with a minor disadvantage for the lowest income bucket. In Brazil, on the other hand, a clear signal emerges, with a monotonically increasing trend of the entropy with income, indicating that the wealthy have access to a much wider mix of services as compare to the poor. This reinforces the advantage of the wealthy when coupled with the observation of closer access to a large fraction of total basic amenities as seen in  Supplementary Fig. S8{\bf D}. Even in the United States, the situation is more nuanced when we examine cities belonging to different clusters. In Fig.~\ref{fig:amenitiesdiversitydetroitbostonriocampinas}{\bf C} we plot the trend for Detroit, a representative city in the teal cluster, and in Fig.~\ref{fig:amenitiesdiversitydetroitbostonriocampinas}{\bf E}, New York, an exemplar of the orange cluster.  In Detroit we once again see a monotonic trend of $H_l^a$ with $ID_{q}$ indicating that while people of lower incomes are proximate to amenities in terms of their number, they have a distinct disadvantage with respect to the wealthy in terms of diversity of such services. In NY, on the other hand, the trend is flat, indicating that there is no income advantage when either accessing the number or type of amenities in the city. Thus similar to the mobility metrics, the primary difference between the the two clusters is the presence or absence of any correlation with income.  

For the case of Brazil, Rio de Janeiro a city from the blue cluster also shows a wealth advantage (Fig.~\ref{fig:amenitiesdiversitydetroitbostonriocampinas}{\bf D}) with a similar trend to Detroit, with a more pronounced difference between the lowest and highest income groups.  Campinas (Fig.~\ref{fig:amenitiesdiversitydetroitbostonriocampinas}{\bf F}) from the yellow cluster, shows more noisy trends, although even here there is a marked difference in the entropy between the lowest and highest income groups. The primary difference between the two clusters is that in one there is a clear monotonic trend, and in the other there is variability in the middle income range.

\subsection{Spatial Effects}

Our analysis thus far has neglected any spatial effects, although the observed trends hint at differences in how cities are organized in terms of residences and their income distribution. For instance, we speculated that the difference between the orange and teal clusters in the United States are reflective of the differences in each city-type in how residences and amenities are spatially located. We next investigate whether such features indeed exist. 

We begin by explicitly measuring the destinations of the upper 20\% and lower 20\% of SA's in terms of median income. For each SA, we examine the in-flow and trace back to the origin of those flows to the income profile of the neighborhoods they originate from. For those regions where the in-flows are predominantly from low income areas, we color the region red, and for those where the origins are from high-income areas we color the region blue. In areas where there is an overlap of both flavors of flow, the regions are colored purple. In the left panel of Fig. \ref{fig:spatial_mixing} we show the results for four American cities, Detroit and Rochester from the teal cluster, and Atlanta and Boston from the orange cluster. For the latter cities, while there are regions exclusively visited by either high- or low-income groups, there are large areas of overlap particularly in the central parts of the city. Conversely, in both Rochester and Detroit, we see a more segregated profile, where there are two distinct regions of high- and low-income areas with relatively less overlap as compared to cities in the orange cluster. Additionally, as suspected from the results of our clustering analysis, the lower-income group tend to visit the city center, whereas the visits by higher-income groups are more concentrated in suburban areas.
In the right panel of Fig.~\ref{fig:spatial_mixing} we show the case for four Brazilian cities, with Brasilia and Campinas from the yellow cluster, and Sao Paolo and Rio de Janeiro from the blue cluster. In all four cities, there is a clear separation of regions visited in terms of income, with practically no overlap. Furthermore, unlike in the US, visits by the wealthy are overwhelmingly concentrated in the central part of the cities, with the poor by and large traversing the periphery. In terms of mixing of the income groups, there is little difference between the the two Brazilian clusters.

\begin{figure}[!t]
	\centering
	\includegraphics[width=0.7\textwidth]{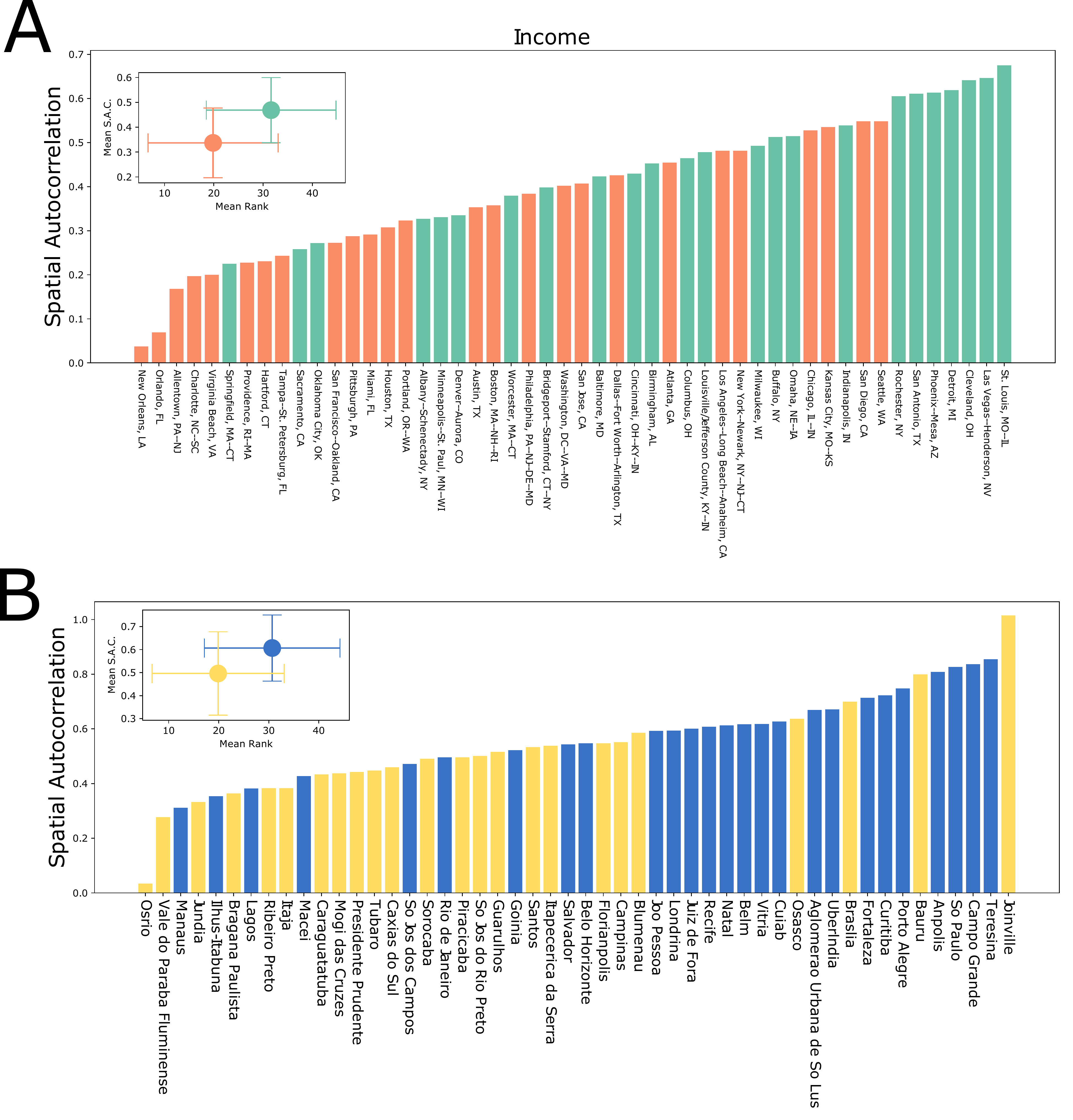}
	\caption{\textbf{Income spatial autocorrelation.} Ranking of Moran's I values for cities in both the US (\textbf{A}) and Brazil (\textbf{B}) colored according to the cluster assigned. A high value ($\sim 1$) indicates that zip codes are grouped closely with zip codes of similar income, while a value close to $0$ indicates a random arrangement. The average values for each cluster are shown as inset.}
\label{fig:spatial_autocorr}
\end{figure}

Where people visit in a city are reflective of the residential patterns. To determine how these are distributed, we next compute the spatial autocorrelation or Moran's I $(I_M$) of the SAs in each city in terms of the median income in that area (see Supplementary Information Section S7 for details of the calculation). The spatial autocorrelation lies in the range $-1 \leq I_M \leq 1$, and is a measure for the similarity between the incomes of adjacent SAs. Its maximum value occurs for a perfectly segregated arrangement where all high- and low-income neighborhoods are adjacent only to each other, whereas the lowest values occur for a perfectly uniform distribution of incomes. A random arrangement with no spatial correlation yields values close to zero.

In Fig.~\ref{fig:spatial_autocorr}{\bf A} we plot the results for the 50 cities in the United States, with each city colored according to the cluster it belongs to. As an inset we show the average values $\langle I_M \rangle$ for each cluster. A very clear trend emerges where cities in the teal cluster have markedly higher values of $\langle I_M \rangle$ than those in the orange cluster, indicating a more segregated profile of neighborhoods in terms of income. This serves as a quantitative confirmation for the different trends seen in each cluster in terms of their correlation with income and patterns of access to the number and diversity of amenities. For cities in the orange cluster, neighborhoods are organized in a more random fashion than the clustered ones in the teal cluster, and provides a possible explanation for the relatively more equitable access to basic services, the higher mixing in visitation patterns, and the general lack of correlations of mobility with income indicators. The comparatively more segregated residential patterns provide one possible causative mechanism for the trends seen in the teal cluster.  

In Fig.~\ref{fig:spatial_autocorr}{\bf B} we show the results for Brazil, once again finding a clear separation between the blue and yellow clusters, with the former having a larger value of $\langle I_M\rangle$. While the differences in the connection between mobility and income is less pronounced between the two clusters, we recall that the blue cluster had higher correlations than the yellow one. The fact that cities in the blue cluster are more segregated, provides further evidence for residential patterns being a causative mechanism for the observed mobility patterns. 

\subsection{Limitations}
These results should be interpreted in light of important limitations. First, the Google mobility data is limited to smartphone users who have opted in to Google's Location History feature. These data may not be representative of the population as whole in all parts of the world, and their representativeness may vary by location. Importantly, these limited data are only viewed through the lens of differential privacy algorithms, specifically designed to protect user anonymity and obscure fine detail. 
Limitations also apply in terms of the data on amenities, given the nature of the source (based on voluntary reporting).

\section{Discussion}

Taken together our results shed greater clarity on the connection between socioeconomic features and movement patterns as compared to previous studies, given the scale of our investigation. A study of a hundred cities in two large countries comparable in size and population but differing in levels of economic development, revealed that cities can be broadly classified into two categories; those where resident's movements are influenced by their income profile (with varying degrees of influence), and those where mobility is largely independent of their socioeconomic condition. This classification of course applies in an averaged sense, given the differences in cities both between and within the countries. 

The two categories find their clearest manifestation in the United States, with a roughly even split between the 50 cities, that was uncovered with a clustering approach, with cities in the income-correlated cluster labeled teal and those in the weakly-correlated cluster, orange. The wider availability of public transportation in the orange cluster appears to be a feature that mitigates the correlation between income and movement, perhaps playing the role of an equalizing feature for those in the lower income brackets. Another key aspect is the availability and diversity of basic services. While lower-income residents in the teal cluster are on average closer to basic amenities, this appears to be a function of them living in the city-center, with the higher-income groups concentrated in suburbs. When it comes to the diversity of services available, there is a clear advantage to those in the higher-income bracket. These cities are also segregated by income, whereby high-income and low-income residents tend to be concentrated in specific areas. This is also manifested in the parts of the cities that they visit, with relatively little mixing between the income groups. Indeed, given the connection with income and race (Figs. S1, S2) this is also a reflection of racial segregation.  By contrast, residents in cities belonging to the orange cluster have comparatively more equitable access to services in terms of both proximity and quality, and are less segregated in terms of both where they live and where they choose to visit.

In Brazil, we also find two categories, however unlike in the United States, movement in both clusters is correlated with socioeconomic features, with the difference being one cluster (blue) displays even stronger correlations then the other (yellow). While public transportation usage is comparable across cities, the accessibility to services in terms of proximity and quality is strongly skewed towards the highest income brackets in the blue cluster, although the income advantage is also found to a lesser extent in the yellow cluster. Additionally cities in the blue cluster have a more segregated residential profile than those in the yellow one.  Both clusters, however display, a sharp divide between rich and poor in terms of which parts of the city is visited, with the rich being concentrated in the city-center and the poor primarily moving in the periphery. In some sense, the Brazilian cities mirror the teal cluster from the United States, with stronger differences in mobility in terms of the socioeconomic characteristics. 

To validate our results, we also conducted our analysis on a different dataset; the commuting patterns extracted from the United States census. The qualitative results are essentially identical, with some differences in the clustering of the cities, stemming from the fact that the census does not accurately capture the frequency of travel. This resulted in a few cities changing clusters, but by and large the differences in terms of transportation usage, accessibility of services and residential segregation were preserved, attesting to the robustness of our analysis. A similar analysis could not be conducted for Brazil given that such data is not available in that country's census. This points to another strength of this new approach: its applicability globally by leveraging aggregate signals such as mobility that are inexpensive to compute and available in a timely fashion. Methods relying on traditional census data do not generalize to many regions where census is unavailable, unreliable, or considerably delayed in time.

To the extent that it is undesirable to have cities with residents whose ability to navigate and access resources is dependent on their socioeconomic status, public policy measures to mitigate this phenomenon are the need of the hour. While it is difficult to disentangle the causal mechanisms behind the observed disparities, investment in affordable and robust public transportation may enhance equality of access~\cite{Teunissen_2015}, although the extent of its efficacy may vary from region to region, given the observed differences between the United States and Brazil. Certainly more efforts must be made in terms of improving the quality of services in low-income areas. Given that this disparity might arise from the residential patterns seen in both countries, inclusive zoning and incentives to foster mixed-income neighborhoods may well be an effective policy~\cite{Graves_2010, Joseph_2010}. Indeed, such initiatives have been tried in cities such as New York, Boston and Chicago, which all incidentally lie in the orange cluster.


Nevertheless, much remains to be uncovered, in particular whether the reported trends here exist in countries in different regions and varying degrees of economic development. Effective strategies and measures will also vary from region to region and more tailored studies would be required to identify the contextual causes of different socio-mobility profiles. We anticipate many more studies along this direction in the near future.


\end{document}


\renewcommand{\thefigure}{S\arabic{figure}}
\renewcommand{\thesection}{S\arabic{section}}
\renewcommand{\thetable}{S\arabic{table}}
\renewcommand{\theequation}{S\arabic{equation}}
\renewcommand{\thepage}{S-\arabic{page}}
\makeatletter
\renewcommand{\maketitle}{\bgroup\setlength{\parindent}{0pt}
	\begin{flushleft}
{\huge Supporting Information}\\[0.3cm]
		{\large\textbf \@title}
		
		\@author
		
	\end{flushleft}\egroup
}
\makeatother
\title{Uncovering the socioeconomic facets of human mobility}

\newcommand{\floor}[1]{\lfloor #1 \rfloor}

\author{Hugo Barbosa, Surendra Hazarie, Brian Dickinson, Aleix Bassolas, Adam Frank, Henry Kautz, Adam Sadilek, Jos\'e J. Ramasco, and Gourab Ghoshal}               

\date{}

\maketitle


%

\section{Census data and income metrics}
\begin{figure}[ht!]
    \centering
    \includegraphics[width=0.7\textwidth]{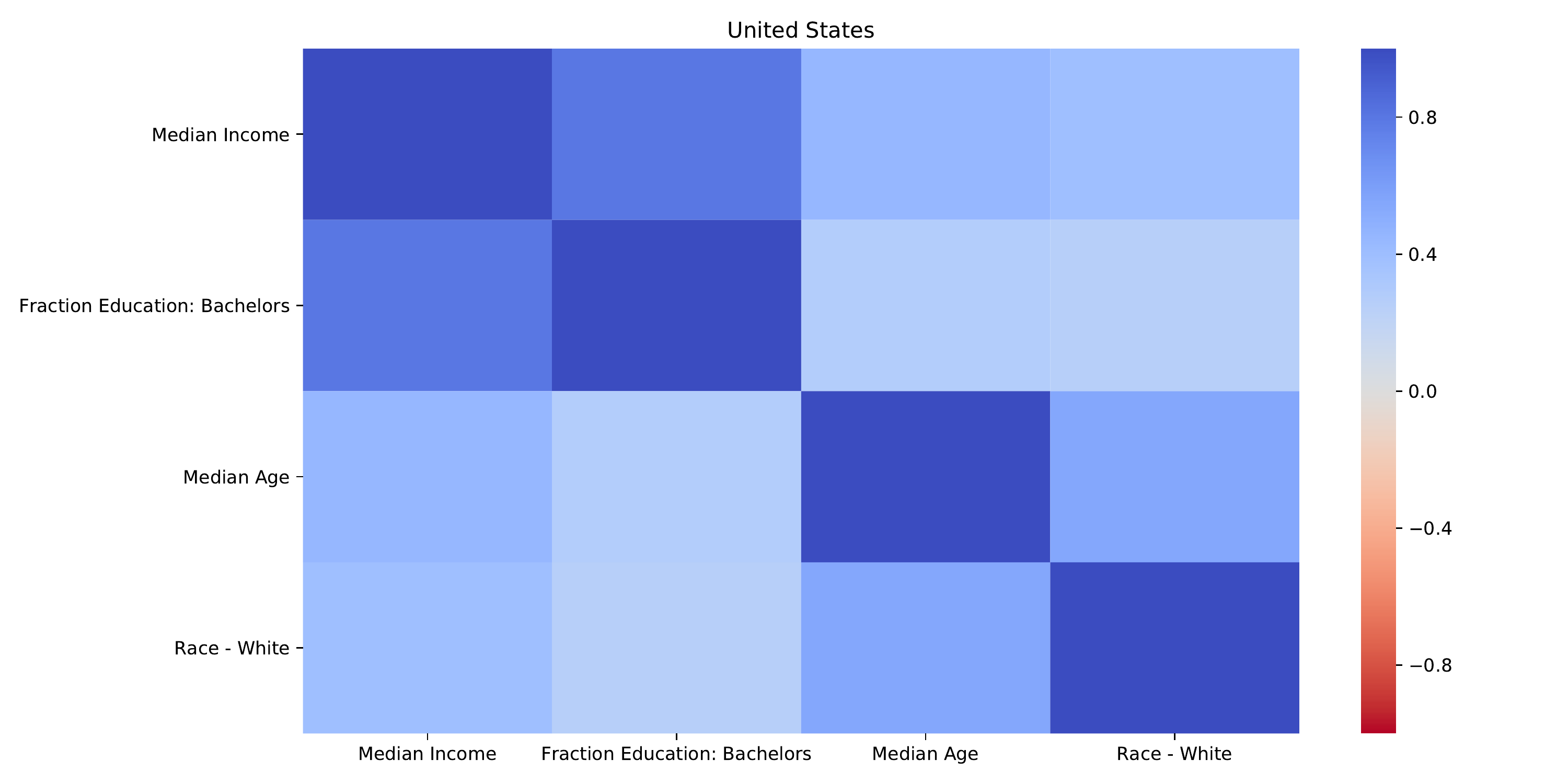}
    \caption{Spearman correlations between 4 selected dimensions of the survey data taken from 50 American cities.}
    \label{fig:us_corr}
\end{figure}

\begin{figure}[ht!]
    \centering
    \includegraphics[width=0.7\textwidth]{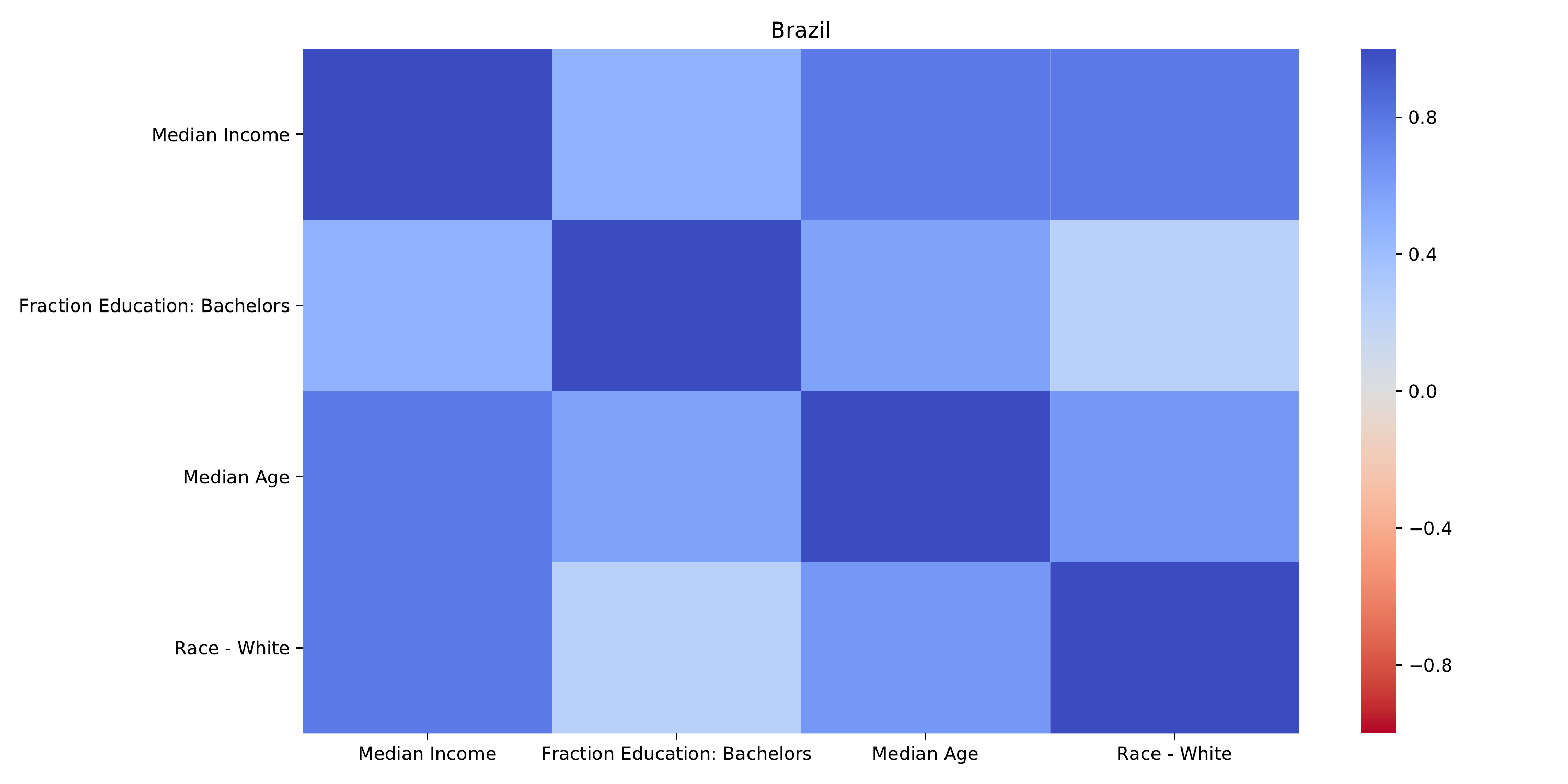}
    \caption{Spearman correlations between 4 selected dimensions of the survey data taken from 50 Brazilian cities.}
    \label{fig:br_corr}
\end{figure}

To represent socioeconomic status we collected information on median income, age, sex composition and race from the 2016 5-year American Community Survey (ACS) estimates for the United States and the 2010 decennial census for Brazil [38,39].  The full set of cities are shown in Tabs.~\ref{tab:usa_selected_cities} and ~\ref{tab:br_selected_cities}. An example of the data for the city of Rochester, NY is shown in Tab.~\ref{tab:socioeconomic_vars}, and for Campinas, Brazil in Tab.~\ref{tab:socioeconomic_vars_2}. The spearman correlation between these variables indicates a strong correlation with median income in both the United States (Fig. \ref{fig:us_corr}) and Brazil (Fig. \ref{fig:br_corr}).

To capture income, we utilize two different representations of the income data, both obtained from the nominal income household variables. The first is the fraction $f_{i}$ of the populations within an area having a per capita income within one of the $n$ income brackets, as defined by the corresponding census authorities, such that $\sum_{i = 1}^{n} f_{1} = 1$. For the case of the US, there were $n=8$ income breaks, ranging from below 10k to above 75kUSD per year. Therefore, each area in the data is characterized by a vector of $n$ values. For instance, if in a given area in the US data (e.g., ZCTA) has an income breakdown vector \[I=[0.09, 0.1, 0.15, 0.15, 0.25, 0.15, 0.10, 0.01],\] it means that 9 \% of that population has a income per-capita below 10k USD a year whereas 1 \% has an income above 75k USD. We refer to this fraction-based income variable as $ID_{f}$.

Similarly for Brazil,  household-level incomes are partitioned into $n=9$ intervals relative to the Brazilian National Minimum Wage (NMW), ranging from less than $1/2$ NMW to nominal incomes above 20 NMWs. 
The second variable is extracted from the median income per capita of an area (ZCTA or WA). From this median income, we split the regions into five income quantiles, which we label as \textit{low} (bottom 20\%), \textit{mid-low} (between 20\%-40\%), \textit{mid} (between 40\%-60\%), \textit{mid-high} (between 60\%-80\%) and \textit{high} (above 80\%).  These income quantiles are relative to each individual city, accounting therefore for the socioeconomic differences across cities. This metric, referred to as  $ID_{q}$ allows us to label an area of a city according to its overall income characteristics.

\begin{table}[!htbp]
    \centering
  \scalebox{0.8}{
\begin{tabular}{llr}
\toprule
{} &                                         Urban Area &  Population \\
\midrule
1  &             Albany--Schenectady, NY Urbanized Area &      669473 \\
2  &                   Allentown, PA--NJ Urbanized Area &      822617 \\
3  &                         Atlanta, GA Urbanized Area &     5381110 \\
4  &                          Austin, TX Urbanized Area &     1711701 \\
5  &                       Baltimore, MD Urbanized Area &     2294493 \\
6  &                      Birmingham, AL Urbanized Area &      855126 \\
7  &                  Boston, MA--NH--RI Urbanized Area &     4704737 \\
8  &        Bridgeport--Stamford, CT--NY Urbanized Area &     1019263 \\
9  &                         Buffalo, NY Urbanized Area &     1038718 \\
10 &                   Charlotte, NC--SC Urbanized Area &     1747070 \\
11 &                     Chicago, IL--IN Urbanized Area &     8917324 \\
12 &              Cincinnati, OH--KY--IN Urbanized Area &     1848639 \\
13 &                       Cleveland, OH Urbanized Area &     1999159 \\
14 &                        Columbus, OH Urbanized Area &     1512446 \\
15 &   Dallas--Fort Worth--Arlington, TX Urbanized Area &     5996274 \\
16 &                  Denver--Aurora, CO Urbanized Area &     2681764 \\
17 &                         Detroit, MI Urbanized Area &     4035863 \\
18 &                        Hartford, CT Urbanized Area &     1105969 \\
19 &                         Houston, TX Urbanized Area &     5412762 \\
20 &                    Indianapolis, IN Urbanized Area &     1690182 \\
21 &                 Kansas City, MO--KS Urbanized Area &     1645871 \\
22 &            Las Vegas--Henderson, NV Urbanized Area &     1905136 \\
23 &  Los Angeles--Long Beach--Anaheim, CA Urbanized... &    12686477 \\
24 &  Louisville/Jefferson County, KY--IN Urbanized ... &     1078499 \\
25 &                           Miami, FL Urbanized Area &     5832060 \\
26 &                       Milwaukee, WI Urbanized Area &     1486454 \\
27 &       Minneapolis--St. Paul, MN--WI Urbanized Area &     2933169 \\
28 &                     New Orleans, LA Urbanized Area &      967547 \\
29 &        New York--Newark, NY--NJ--CT Urbanized Area &    18987636 \\
30 &                   Oklahoma City, OK Urbanized Area &     1125647 \\
31 &                       Omaha, NE--IA Urbanized Area &      803087 \\
32 &                         Orlando, FL Urbanized Area &     1866002 \\
33 &        Philadelphia, PA--NJ--DE--MD Urbanized Area &     5821389 \\
34 &                   Phoenix--Mesa, AZ Urbanized Area &     4038319 \\
35 &                      Pittsburgh, PA Urbanized Area &     1955719 \\
36 &                    Portland, OR--WA Urbanized Area &     2080902 \\
37 &                  Providence, RI--MA Urbanized Area &     1188589 \\
38 &                       Rochester, NY Urbanized Area &      806269 \\
39 &                      Sacramento, CA Urbanized Area &     1889988 \\
40 &                     San Antonio, TX Urbanized Area &     2091059 \\
41 &                       San Diego, CA Urbanized Area &     3097903 \\
42 &          San Francisco--Oakland, CA Urbanized Area &     3508790 \\
43 &                        San Jose, CA Urbanized Area &     1733914 \\
44 &                         Seattle, WA Urbanized Area &     3429707 \\
45 &                 Springfield, MA--CT Urbanized Area &      674903 \\
46 &                   St. Louis, MO--IL Urbanized Area &     2309620 \\
47 &           Tampa--St. Petersburg, FL Urbanized Area &     2689944 \\
48 &                  Virginia Beach, VA Urbanized Area &     1554524 \\
49 &              Washington, DC--VA--MD Urbanized Area &     5098266 \\
50 &                   Worcester, MA--CT Urbanized Area &      541879 \\
\bottomrule
\end{tabular}%
}

    \caption{List of US cities and their populations}
    \label{tab:usa_selected_cities}
\end{table}

\begin{table}[!htbp]
    \centering
    \scalebox{0.8}{
\begin{tabular}{llr}
\toprule
{} &                  Urban Area &  Population \\
\midrule
1  &  Aglomerao Urbana de So Lus &     1014741 \\
2  &                     Anpolis &      332727 \\
3  &                       Bauru &      343877 \\
4  &                        Belm &     1865652 \\
5  &              Belo Horizonte &     3854722 \\
6  &                    Blumenau &      308402 \\
7  &            Bragana Paulista &      160665 \\
8  &                     Braslia &     2566723 \\
9  &                    Campinas &     1924733 \\
10 &                Campo Grande &      785683 \\
11 &               Caraguatatuba &      113317 \\
12 &               Caxias do Sul &      435791 \\
13 &                       Cuiab &      803497 \\
14 &                    Curitiba &     2227447 \\
15 &                Florianpolis &      629936 \\
16 &                   Fortaleza &     2985515 \\
17 &                      Goinia &     1757432 \\
18 &                   Guarulhos &     1222308 \\
19 &               Ilhus-Itabuna &      204310 \\
20 &                       Itaja &        5544 \\
21 &        Itapecerica da Serra &      714128 \\
22 &                   Joinville &      515128 \\
23 &                  Joo Pessoa &      722866 \\
24 &                Juiz de Fora &      516417 \\
25 &                      Jundia &      369862 \\
26 &                       Lagos &    14368000 \\
27 &                    Londrina &      507114 \\
28 &                       Macei &      931636 \\
29 &                      Manaus &     1793000 \\
30 &             Mogi das Cruzes &      968510 \\
31 &                       Natal &     1006137 \\
32 &                      Osasco &     1474647 \\
33 &                       Osrio &       40000 \\
34 &                  Piracicaba &      363890 \\
35 &                Porto Alegre &     2876717 \\
36 &         Presidente Prudente &      207841 \\
37 &                      Recife &     2808854 \\
38 &               Ribeiro Preto &      604494 \\
39 &              Rio de Janeiro &    10816739 \\
40 &                    Salvador &     2917528 \\
41 &                      Santos &     1297281 \\
42 &         So Jos do Rio Preto &      408079 \\
43 &           So Jos dos Campos &     1118953 \\
44 &                    So Paulo &    12698814 \\
45 &                    Sorocaba &      586084 \\
46 &                    Teresina &      812594 \\
47 &                      Tubaro &      104937 \\
48 &                   Uberlndia &      604148 \\
49 &   Vale do Paraba Fluminense &      232296 \\
50 &                      Vitria &     1497160 \\
\bottomrule
\end{tabular}

}

    \caption{List of BR cities and their populations}
    \label{tab:br_selected_cities}
\end{table}
\clearpage


\begin{table}[]
\scalebox{0.8}{
\begin{tabular}{ccccccc}
\toprule
{} &     Urban Area &   ZCTA &  Median Income (\$) &  Fraction Bachelors degree  &  Median Age &  Fraction White \\
\midrule
8429 &  Rochester, NY &  14424 &            29692.0 &                          0.177 &        46.3 &                 0.948 \\
8430 &  Rochester, NY &  14425 &            34762.0 &                          0.264 &        37.0 &                 0.913 \\
8431 &  Rochester, NY &  14428 &            32452.0 &                          0.209 &        47.1 &                 0.977 \\
8432 &  Rochester, NY &  14445 &            31355.0 &                          0.196 &        35.9 &                 0.883 \\
8433 &  Rochester, NY &  14450 &            39894.0 &                          0.302 &        45.3 &                 0.930 \\
8434 &  Rochester, NY &  14467 &            32915.0 &                          0.189 &        39.8 &                 0.770 \\
8435 &  Rochester, NY &  14468 &            32431.0 &                          0.163 &        42.7 &                 0.970 \\
8436 &  Rochester, NY &  14502 &            33661.0 &                          0.197 &        41.2 &                 0.963 \\
8437 &  Rochester, NY &  14506 &            59107.0 &                          0.355 &        45.9 &                 0.986 \\
8438 &  Rochester, NY &  14514 &            33598.0 &                          0.208 &        41.3 &                 0.890 \\
8439 &  Rochester, NY &  14519 &            30990.0 &                          0.160 &        44.6 &                 0.948 \\
8440 &  Rochester, NY &  14526 &            40069.0 &                          0.258 &        44.5 &                 0.919 \\
8441 &  Rochester, NY &  14534 &            47629.0 &                          0.328 &        46.1 &                 0.888 \\
8442 &  Rochester, NY &  14543 &            36729.0 &                          0.242 &        50.4 &                 0.965 \\
8443 &  Rochester, NY &  14544 &            28083.0 &                          0.102 &        43.1 &                 0.988 \\
8444 &  Rochester, NY &  14546 &            31126.0 &                          0.191 &        41.8 &                 0.874 \\
8445 &  Rochester, NY &  14548 &            30110.0 &                          0.114 &        48.3 &                 0.973 \\
8446 &  Rochester, NY &  14559 &            34586.0 &                          0.229 &        42.9 &                 0.954 \\
8447 &  Rochester, NY &  14564 &            40391.0 &                          0.270 &        42.5 &                 0.954 \\
8448 &  Rochester, NY &  14568 &            32311.0 &                          0.244 &        37.9 &                 0.983 \\
8449 &  Rochester, NY &  14580 &            35109.0 &                          0.250 &        43.5 &                 0.913 \\
8450 &  Rochester, NY &  14586 &            36719.0 &                          0.255 &        32.9 &                 0.756 \\
8451 &  Rochester, NY &  14589 &            27301.0 &                          0.124 &        45.3 &                 0.928 \\
8452 &  Rochester, NY &  14604 &            16330.0 &                          0.083 &        43.9 &                 0.550 \\
8453 &  Rochester, NY &  14605 &            11290.0 &                          0.073 &        25.8 &                 0.285 \\
8454 &  Rochester, NY &  14606 &            23630.0 &                          0.109 &        36.4 &                 0.631 \\
8455 &  Rochester, NY &  14607 &            27811.0 &                          0.320 &        30.0 &                 0.799 \\
8456 &  Rochester, NY &  14608 &            15074.0 &                          0.128 &        29.6 &                 0.225 \\
8457 &  Rochester, NY &  14609 &            25608.0 &                          0.146 &        33.5 &                 0.544 \\
8458 &  Rochester, NY &  14610 &            39575.0 &                          0.266 &        44.1 &                 0.875 \\
8459 &  Rochester, NY &  14611 &            16054.0 &                          0.062 &        30.7 &                 0.267 \\
8460 &  Rochester, NY &  14612 &            31247.0 &                          0.201 &        45.5 &                 0.894 \\
8461 &  Rochester, NY &  14613 &            19657.0 &                          0.097 &        30.5 &                 0.400 \\
8462 &  Rochester, NY &  14614 &             8894.0 &                          0.041 &        28.7 &                 0.360 \\
8463 &  Rochester, NY &  14615 &            23572.0 &                          0.148 &        35.9 &                 0.634 \\
8464 &  Rochester, NY &  14616 &            28127.0 &                          0.159 &        41.5 &                 0.847 \\
8465 &  Rochester, NY &  14617 &            34438.0 &                          0.272 &        44.2 &                 0.930 \\
8466 &  Rochester, NY &  14618 &            36906.0 &                          0.274 &        40.1 &                 0.859 \\
8467 &  Rochester, NY &  14619 &            22250.0 &                          0.114 &        32.6 &                 0.199 \\
8468 &  Rochester, NY &  14620 &            24646.0 &                          0.252 &        31.4 &                 0.714 \\
8469 &  Rochester, NY &  14621 &            15884.0 &                          0.060 &        33.0 &                 0.343 \\
8470 &  Rochester, NY &  14622 &            30134.0 &                          0.176 &        46.4 &                 0.881 \\
8471 &  Rochester, NY &  14623 &            21566.0 &                          0.215 &        27.7 &                 0.722 \\
8472 &  Rochester, NY &  14624 &            30507.0 &                          0.204 &        41.2 &                 0.838 \\
8473 &  Rochester, NY &  14625 &            36157.0 &                          0.272 &        48.4 &                 0.956 \\
8474 &  Rochester, NY &  14626 &            30845.0 &                          0.176 &        47.0 &                 0.886 \\
\bottomrule
\end{tabular}
}
\caption{Socioeconomic variables for zip codes in Rochester, NY}
\label{tab:socioeconomic_vars}

\end{table}

\begin{table}[]
\scalebox{0.8}{
\begin{tabular}{ccccccc}
\toprule
{} & Urban Area &      Area Code &  Median Income (\$) &  Fraction Bachelors degree &  Median Age &  Fraction White\\
\midrule
1287 &   Campinas &  3501608005001 &             1500.0 &                          0.278 &        41.0 &                 0.913 \\
1288 &   Campinas &  3501608005002 &             1000.0 &                          0.255 &        33.0 &                 0.714 \\
1289 &   Campinas &  3501608005003 &              850.0 &                          0.209 &        30.0 &                 0.698 \\
1290 &   Campinas &  3501608005004 &             1200.0 &                          0.262 &        33.0 &                 0.791 \\
1291 &   Campinas &  3501608005005 &             1500.0 &                          0.270 &        33.0 &                 0.795 \\
1292 &   Campinas &  3501608005006 &             1000.0 &                          0.249 &        31.0 &                 0.733 \\
1293 &   Campinas &  3501608005007 &             1400.0 &                          0.231 &        36.0 &                 0.815 \\
1294 &   Campinas &  3501608005008 &             1500.0 &                          0.295 &        38.0 &                 0.907 \\
1295 &   Campinas &  3501608005009 &             1500.0 &                          0.290 &        38.0 &                 0.862 \\
1296 &   Campinas &  3501608005010 &             1000.0 &                          0.252 &        32.0 &                 0.743 \\
1342 &   Campinas &  3509502005001 &              940.0 &                          0.173 &        33.0 &                 0.657 \\
1343 &   Campinas &  3509502005002 &             1500.0 &                          0.249 &        36.0 &                 0.802 \\
1344 &   Campinas &  3509502005003 &             1750.0 &                          0.258 &        32.0 &                 0.770 \\
1345 &   Campinas &  3509502005004 &             2500.0 &                          0.221 &        34.0 &                 0.812 \\
1346 &   Campinas &  3509502005005 &             2000.0 &                          0.266 &        39.0 &                 0.782 \\
1347 &   Campinas &  3509502005006 &             2000.0 &                          0.259 &        34.0 &                 0.804 \\
1348 &   Campinas &  3509502005007 &             1850.0 &                          0.283 &        37.0 &                 0.860 \\
1349 &   Campinas &  3509502005008 &             1200.0 &                          0.228 &        34.0 &                 0.743 \\
1350 &   Campinas &  3509502005009 &             1500.0 &                          0.290 &        36.0 &                 0.787 \\
1351 &   Campinas &  3509502005010 &             2500.0 &                          0.261 &        37.0 &                 0.874 \\
1352 &   Campinas &  3509502005011 &             2850.0 &                          0.187 &        35.0 &                 0.836 \\
1353 &   Campinas &  3509502005012 &             4200.0 &                          0.216 &        42.0 &                 0.906 \\
1354 &   Campinas &  3509502005013 &             1700.0 &                          0.325 &        36.0 &                 0.799 \\
1355 &   Campinas &  3509502005014 &             2000.0 &                          0.281 &        39.0 &                 0.897 \\
1356 &   Campinas &  3509502005015 &             1110.0 &                          0.263 &        32.0 &                 0.716 \\
1357 &   Campinas &  3509502005016 &             1200.0 &                          0.291 &        37.0 &                 0.747 \\
1358 &   Campinas &  3509502005017 &             1000.0 &                          0.245 &        31.0 &                 0.622 \\
1359 &   Campinas &  3509502005018 &             1000.0 &                          0.262 &        31.0 &                 0.641 \\
1360 &   Campinas &  3509502005019 &             1500.0 &                          0.269 &        33.0 &                 0.712 \\
1361 &   Campinas &  3509502005020 &             1510.0 &                          0.310 &        37.0 &                 0.793 \\
1362 &   Campinas &  3509502005021 &             1000.0 &                          0.247 &        31.0 &                 0.651 \\
1363 &   Campinas &  3509502005022 &             1500.0 &                          0.272 &        35.0 &                 0.787 \\
1364 &   Campinas &  3509502005023 &             1200.0 &                          0.270 &        34.0 &                 0.724 \\
1365 &   Campinas &  3509502005024 &              800.0 &                          0.170 &        26.0 &                 0.457 \\
1366 &   Campinas &  3509502005025 &              900.0 &                          0.239 &        28.0 &                 0.499 \\
1367 &   Campinas &  3509502005026 &              826.4 &                          0.192 &        27.0 &                 0.499 \\
1368 &   Campinas &  3509502005027 &              900.0 &                          0.200 &        27.0 &                 0.470 \\
1369 &   Campinas &  3509502005028 &             1000.0 &                          0.273 &        30.0 &                 0.512 \\
1370 &   Campinas &  3509502005029 &              860.0 &                          0.192 &        27.0 &                 0.484 \\
1371 &   Campinas &  3509502005030 &              900.0 &                          0.205 &        28.0 &                 0.484 \\
1372 &   Campinas &  3509502005031 &              850.0 &                          0.208 &        29.0 &                 0.512 \\
1373 &   Campinas &  3509502005032 &              700.0 &                          0.134 &        24.0 &                 0.378 \\
1374 &   Campinas &  3509502005033 &              930.0 &                          0.226 &        28.0 &                 0.585 \\
1375 &   Campinas &  3509502005034 &             1000.0 &                          0.229 &        31.0 &                 0.572 \\
1376 &   Campinas &  3509502005035 &             1100.0 &                          0.271 &        31.0 &                 0.659 \\
1377 &   Campinas &  3509502005036 &              800.0 &                          0.195 &        27.0 &                 0.458 \\
1495 &   Campinas &  3519071005001 &             1200.0 &                          0.296 &        31.0 &                 0.680 \\
1496 &   Campinas &  3519071005002 &             1000.0 &                          0.280 &        30.0 &                 0.549 \\
1497 &   Campinas &  3519071005003 &             1000.0 &                          0.272 &        30.0 &                 0.608 \\
1498 &   Campinas &  3519071005004 &              839.0 &                          0.191 &        27.0 &                 0.475 \\
\bottomrule
\end{tabular}
}
\caption{Socioeconomic variables for area codes in Campinas}
\label{tab:socioeconomic_vars_2}

\end{table}

\clearpage

\clearpage

\section{Basic amenities data}

\begin{table}[!htbp]
    \centering
    \scalebox{0.8}{
    \begin{tabular}{lcccc}
    \toprule
         & Atlanta, GA & Boston, MA & 
           Detroit, MI & Rochester, NY \\
           \midrule
        marketplace & 15 & 13 & 9 & 0 \\
        restaurant & 1020 & 1447 & 566 & 182 \\
        food court & 2 & 6 & 0 & 0 \\
        fast food & 534 & 513 & 341 & 64 \\
        hospital & 6 & 13 & 7 & 0 \\
        doctors & 39 & 49 & 77 & 6 \\
        dentist & 23 & 67 & 20 & 4 \\
        clinic & 21 & 18 & 40 & 3 \\
        pharmacy & 107 & 129 & 96 & 16 \\
        kindergarten & 11 & 18 & 4 & 2 \\
        school & 1231 & 1253 & 1253 & 339 \\
        college & 11 & 22 & 3 & 5 \\
        university & 4 & 18 & 3 & 1 \\
        library & 84 & 648 & 41 & 36 \\
        bank & 124 & 212 & 103 & 44 \\
        atm & 234 & 164 & 36 & 27\\
        \bottomrule
    \end{tabular}
    }
    \caption{Number of basic amenities by category for Atlanta, GA, Boston MA, Detroit, MI, and Rochester, NY.}
    \label{tab:amenities_example}
\end{table}

\section{Connecting outflow and resident population based on income}
\begin{figure}[h!]
    \centering
    \includegraphics[width=0.7\linewidth]{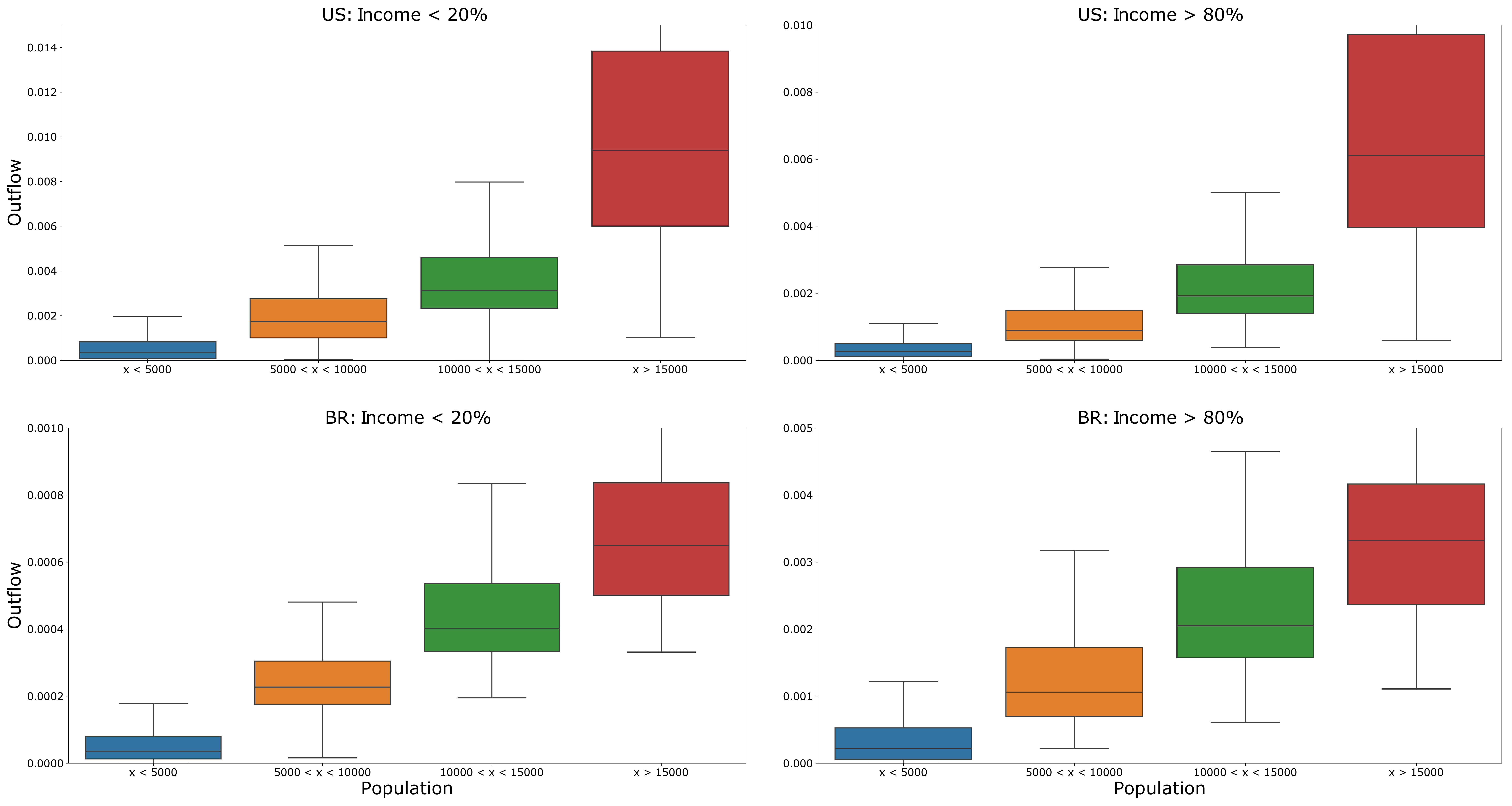}
    \caption{Box-plots of mobility outflows for the top and bottom $20\%$ of zip codes in terms of median income, for the US (top) and Brazil (bottom.) The data is binned by population, and the boxes indicate quartiles of the distributions. We see that in all four cases, population is positively correlated with mobility (Pearson correlations: US $< 20\%: .85, > 80\%: .73$; BR $<20\%: .81, >80\%: .67$) indicating that mobility is representative of population regardless of income.}
    \label{fig:silhouette}
\end{figure}
\clearpage

\section{Cluster analysis}
\label{si:clust}
The clustering of cities was done using a divisive hierarchical method to split the cities into disjoint groups of urban areas with similar patterns in their mobility/socioeconomic correlations. To determine the ideal number of partitions we use a silhouette analysis. 
The silhouette value is a measure of how similar a data point is to the cluster it has been assigned to in comparison to its average distance to another candidate cluster. More precisely, the silhouette score $s(i)$ of a data point $i$ with regards do a partitioning structure $C$ can be defined as 
$$s(i) = \frac{b(i) - a(i) }{\max(a(i),b(i) )}$$
where 
$$a(i) = \frac{1}{|C_{i}| -1} \sum_{j \in C_{i}, i\neq j}  d(i,j)$$
is the mean distance $d(i,j)$ between the point $i$ and all other points ($j$)  in the same cluster $C_{i}$ and 
$$b(i) = \min_{k\neq i} \frac{1}{|C_{k}|}\sum_{j\in C_{k}} d(i,j)$$
where $b(i)$ is the minimum average distance of the point $i$ to the points of one of the other clusters. Here, $b(i)$ is the distance of $i$ to the cluster it is the most similar with, other than the one it has been assigned to. Therefore, $s(i) \in [-1,1]$ such that if the silhouette score $s(i)>0$ that data point is in the best cluster it could belong to. Conversely, if $s(i) < 0 $ the point $i$ is not assigned to its most natural cluster, suggesting that the partitioning structure is not ideal. The ideal partitioning structure is expected to be the one that produces only positive silhouette scores at the same time that maximizes the average silhouette score. Figure~\ref{fig:silhouette} shows the average silhouette scores  for $k=2$ to $k=10$. For both countries $k=2$ produces the maximum average silhouette score. 

\begin{figure}[hbtp]
    \centering
    \includegraphics[width=0.7\linewidth]{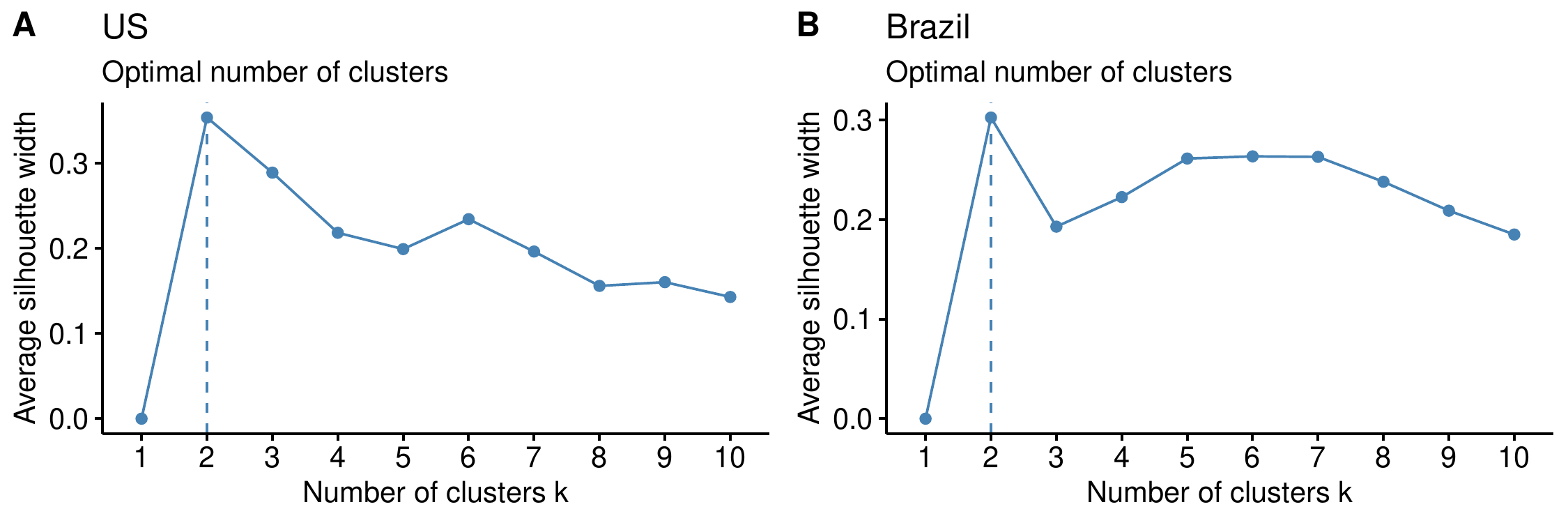}
    \caption{Average Silhouette scores for the partitioning structures produced by the divisive hierarchical clustering algorithm as a function of the number of clusters, $k$ For both the US \textbf{(A)} and Brazil \textbf{(B)} the best partition is obtained for $k=2$.}
    \label{fig:silhouette}
\end{figure}

\clearpage
\section{Commuting data}
\label{sec:si:commuting}

\begin{figure*}[!htbp]
	\centering
	\includegraphics[width=0.4\columnwidth]{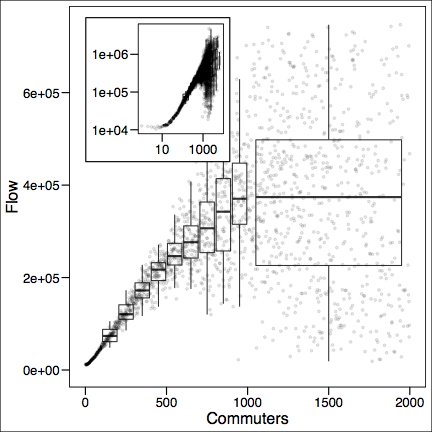}
	\caption{(The mobility flows from the location based data as a function of the commuting flow extracted from the United States census.}
	\label{fig:flow_commuter_correlation}
\end{figure*}

\begin{figure*}[!htbp]
	\centering
	\includegraphics[width=\columnwidth]{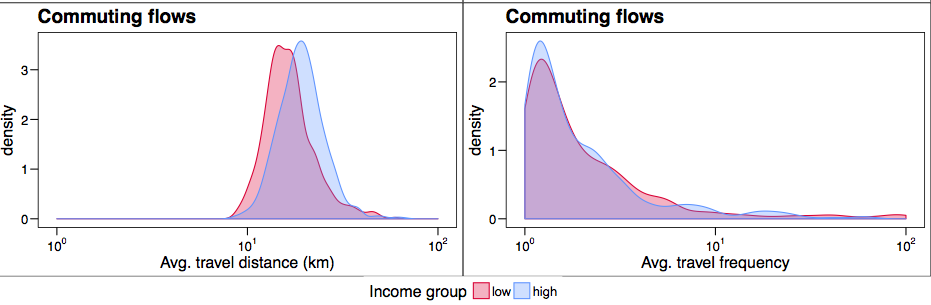}
	\caption{Average travel distance across income groups matches the trends seen in Fig. 2, where higher income groups travel longer distances on average. However, the trip frequency trends are quite different with no discernible difference between groups.}
	\label{fig:flow_commuter_comparison}
\end{figure*}
\clearpage

\begin{figure*}
    \centering
    \includegraphics[width=\textwidth]{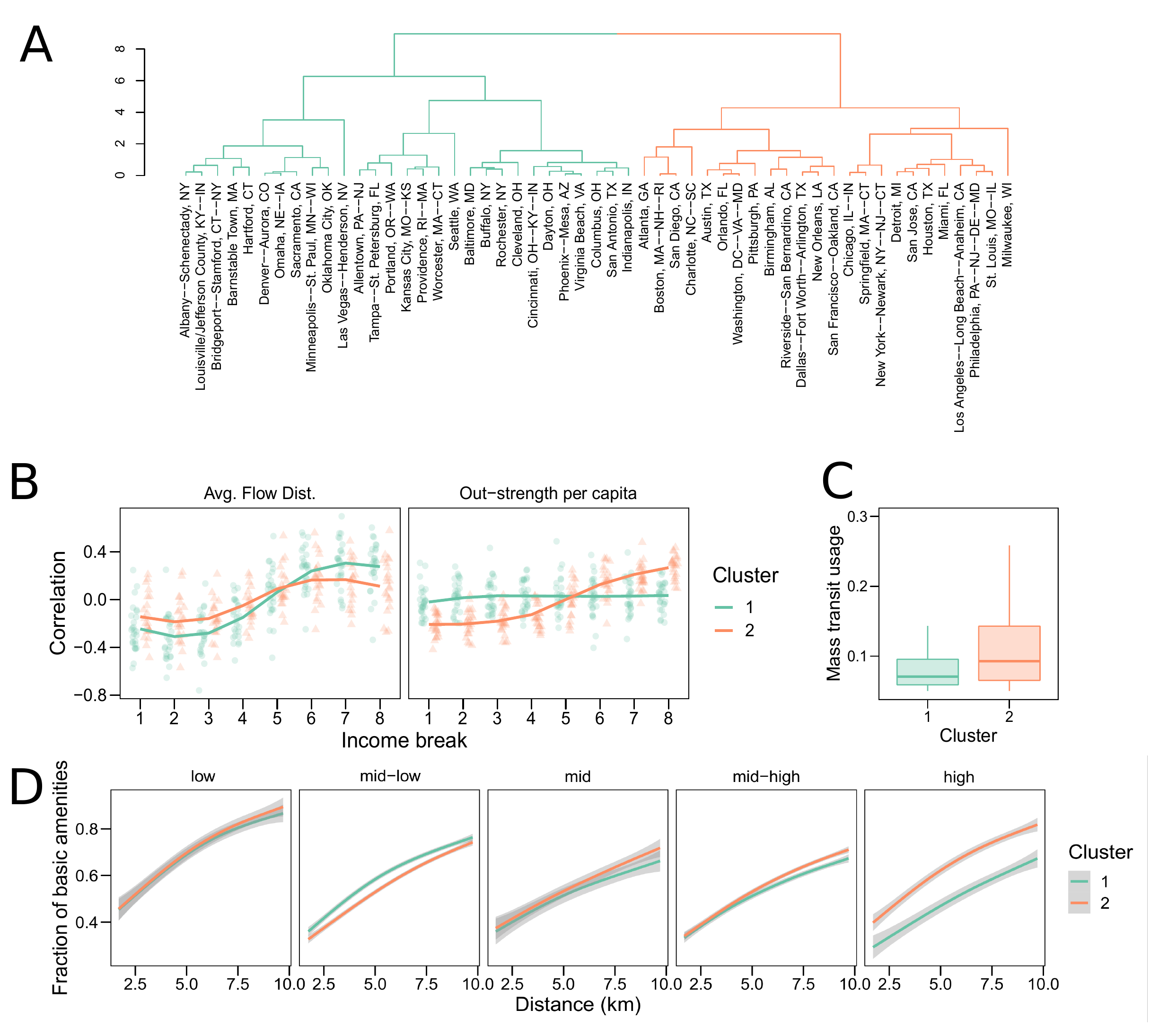}
    \caption{Reproduction of core findings using LODES commuting dataset. {\bf A} The identified city clusters are similar (Fowlkes-Mallows score $=0.8$), though not identical, to the clusters shown in Fig. 4. {\bf B} The trends in the correlation between income and average flow distance are similar to, but less exaggerated than, those found when using location data. The correlation with out-strength per capita however is quite different.  {\bf C} The difference in mass transit usage between clusters {\bf  D} Finally while we see limited differences in amenity distance between clusters at most income levels, we find that in the teal cluster, high-income individuals live on average further from basic amenities, similar to the trend seen in Fig. 4.}
    \label{fig:commuter_clustering_analysis}
\end{figure*}

\clearpage

\begin{figure}[!htbp]
    \centering
    \includegraphics[width=0.7\linewidth]{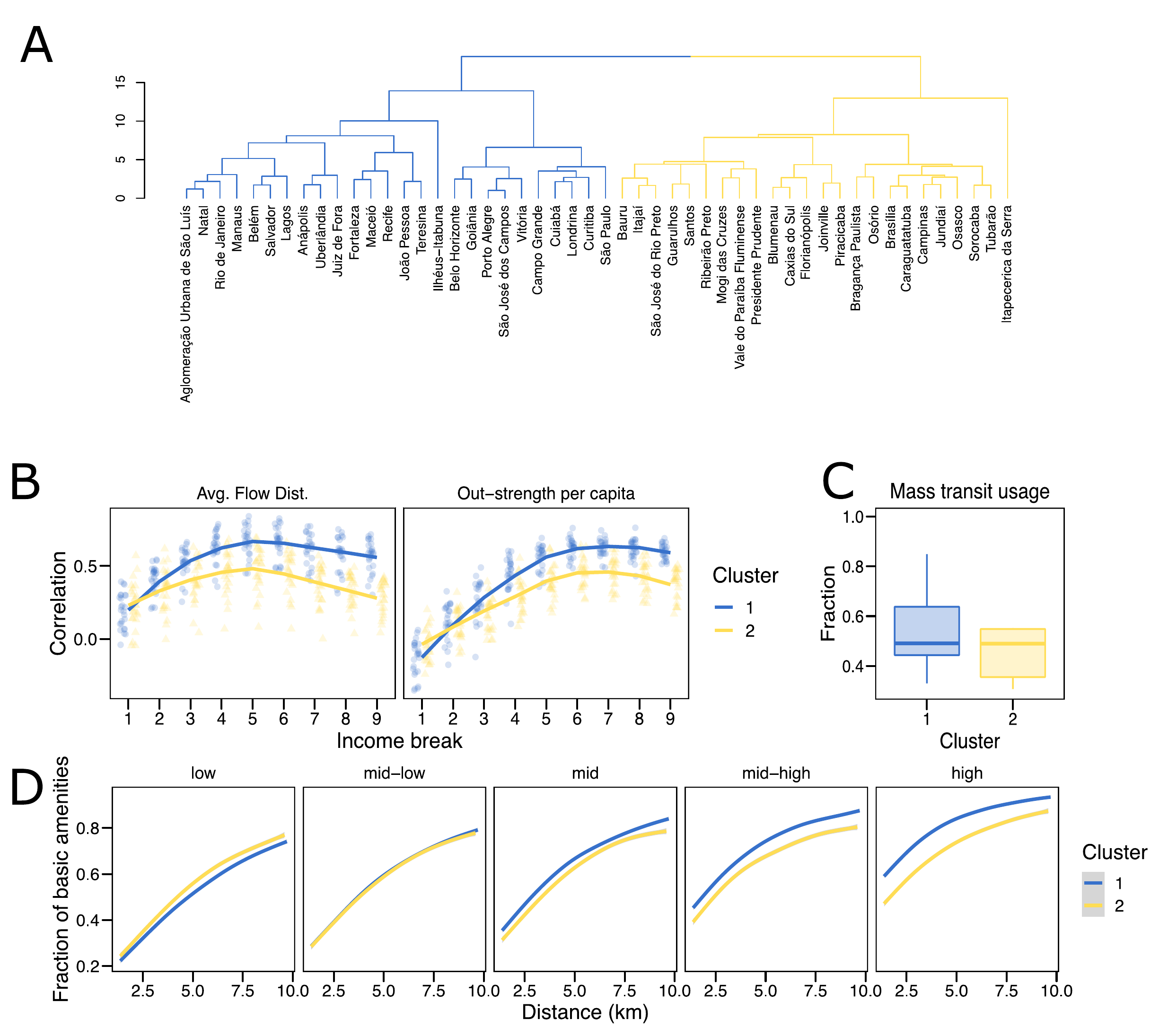}
    \caption{ \textbf{A.} Dendrogram obtained from socio-mobility clustering analysis in Brazilian cities. Cluster 1 (blue) contains almost all the largest metropolitan areas and the capital cities of their respective states, with the exceptions of Lagos and S\~ao Jos\'e dos Campos whereas among those in Cluster 2 (yellow) the only state capital is Florian\'opolis. \textbf{B} Spearman correlation between the number residents of each income bracket and the average flow distance and out-strength per capita. All cities show a high correlation, and the only difference between them is the magnitude of such correlation and how it affects the middle class. \textbf{C} Fraction of mass transit usage for each of the clusters. \textbf{D} Fraction of basic amenities as a function of distance for $5$ income levels.}
    \label{fig:brazil_clusters}
\end{figure}
\clearpage

\section{Amenity diversity}


The amenity diversity was defined as the Shannon entropy, computed from  the relative frequencies of the amenities' distribution. In our analyses we considered a set eight major categories of amenities, namely: $A=$\{\texttt{art-culture,education,entertainment, finance, healthcare, sustenance, transportation, other}\}. For each spatial location unit $l$ we computed the entropy $$
H_l ^a = -\sum_{i\in l} p_{i}\log{p_{i}}
$$
where $p_{i}$ is the fraction of amenities in the area such that $p_{i}= \frac{n_{i}}{N }$ with $N$ being the total number of amenities in the area and $n_{i}$ the number of amenities of type $i$.

\section{Spatial autocorrelation of areas according to income}

The spatial autocorrelation of median incomes across zip codes, is calculated as:

\begin{equation}
    I = \frac{N}{W} \frac{\sum_{i,j}w_{ij} (x_i - \Bar{x}) (x_j - \Bar{x})}{\sum_i(x_i-\Bar{x})^2}
\end{equation}

where $N$ is the total number of zip codes $i$ in a city, $x_i$ is the median income of zip code $i$, $\Bar{x}$ is the mean of all median incomes in the city, $w_{ij} = 1$ if zip code $j$ is touching $i$ and $0$ otherwise, and $W$ is the sum of all $w_{ij}$. 

The metric measures how alike zip codes are to their direct neighbors within a city by comparing them to the city-average. $I$ is near $-1$ for a totally mixed distribution of zip code incomes (zip codes are unlike their neighbors relative to the average, leading to negative numerator contributions), $0$ for a random arrangement, and $1$ if high and low income zip-codes are completely separated (zip codes are completely alike their neighbors, leading to positive numerator contributions.)   We can interpret the autocorrelation of median income as a measure of the degree to which incomes are spatially mixed within a city. Higher values indicate that zip codes of a particular income tend to be grouped together (as in the presence of ``downtown areas",) whereas lower values indicate a more homogeneous distribution throughout a city.

%
%
